%% file: Paper_Final_rev_mar10.tex
\documentclass[aps,prd,twocolumn,superscriptaddress,showpacs,amsmath,amssymb]{revtex4}

\usepackage{graphicx}
\usepackage{dcolumn}
\usepackage{setspace}
\usepackage{subfigure}
\usepackage{slashbox}
\usepackage{epsf}
\usepackage{epsfig}

\begin{document}


\newcommand{\mtop}{\mbox{$M_{\rm top}$}}
\newcommand{\ttbar}{\mbox{$t\bar{t}$}}
\newcommand{\thetastar}{\mbox{$\theta^*$}}
\newcommand{\costhetastar}{\mbox{$\cos\thetastar$}}
\newcommand{\Mlb}{\mbox{$M_{lb}$}}
\newcommand{\Et}{\mbox{$E_T$}}
\newcommand{\Pt}{\mbox{$p_T$}}
\newcommand{\Flong}{\mbox{$F_0$}}
\newcommand{\Fplus}{\mbox{$F_+$}}
\newcommand{\met}{\mbox{$\protect \raisebox{0.3ex}{$\not$}\Et$}}

\bibliographystyle{revtex}

\vspace*{1.5cm}

\title{\boldmath 
        Measurement of the  Top Quark Mass  and $p\bar{p} \to t\bar{t}$
       Cross Section in the All-Hadronic Mode 
       with the CDF\,II Detector}


\input{October2009_Authors.tex}

\input{Paper_Final_body_mar10}

\end{document}

%% file: October2009_Authors.tex
\affiliation{Institute of Physics, Academia Sinica, Taipei, Taiwan 11529, Republic of China} 
\affiliation{Argonne National Laboratory, Argonne, Illinois 60439} 
\affiliation{University of Athens, 157 71 Athens, Greece} 
\affiliation{Institut de Fisica d'Altes Energies, Universitat Autonoma de Barcelona, E-08193, Bellaterra (Barcelona), Spain} 
\affiliation{Baylor University, Waco, Texas  76798} 
\affiliation{Istituto Nazionale di Fisica Nucleare Bologna, $^{dd}$University of Bologna, I-40127 Bologna, Italy} 
\affiliation{Brandeis University, Waltham, Massachusetts 02254} 
\affiliation{University of California, Davis, Davis, California  95616} 
\affiliation{University of California, Los Angeles, Los Angeles, California  90024} 
\affiliation{University of California, San Diego, La Jolla, California  92093} 
\affiliation{University of California, Santa Barbara, Santa Barbara, California 93106} 
\affiliation{Instituto de Fisica de Cantabria, CSIC-University of Cantabria, 39005 Santander, Spain} 
\affiliation{Carnegie Mellon University, Pittsburgh, PA  15213} 
\affiliation{Enrico Fermi Institute, University of Chicago, Chicago, Illinois 60637}
\affiliation{Comenius University, 842 48 Bratislava, Slovakia; Institute of Experimental Physics, 040 01 Kosice, Slovakia} 
\affiliation{Joint Institute for Nuclear Research, RU-141980 Dubna, Russia} 
\affiliation{Duke University, Durham, North Carolina  27708} 
\affiliation{Fermi National Accelerator Laboratory, Batavia, Illinois 60510} 
\affiliation{University of Florida, Gainesville, Florida  32611} 
\affiliation{Laboratori Nazionali di Frascati, Istituto Nazionale di Fisica Nucleare, I-00044 Frascati, Italy} 
\affiliation{University of Geneva, CH-1211 Geneva 4, Switzerland} 
\affiliation{Glasgow University, Glasgow G12 8QQ, United Kingdom} 
\affiliation{Harvard University, Cambridge, Massachusetts 02138} 
\affiliation{Division of High Energy Physics, Department of Physics, University of Helsinki and Helsinki Institute of Physics, FIN-00014, Helsinki, Finland} 
\affiliation{University of Illinois, Urbana, Illinois 61801} 
\affiliation{The Johns Hopkins University, Baltimore, Maryland 21218} 
\affiliation{Institut f\"{u}r Experimentelle Kernphysik, Karlsruhe Institute of Technology, D-76131 Karlsruhe, Germany} 
\affiliation{Center for High Energy Physics: Kyungpook National University, Daegu 702-701, Korea; Seoul National University, Seoul 151-742, Korea; Sungkyunkwan University, Suwon 440-746, Korea; Korea Institute of Science and Technology Information, Daejeon 305-806, Korea; Chonnam National University, Gwangju 500-757, Korea; Chonbuk National University, Jeonju 561-756, Korea} 
\affiliation{Ernest Orlando Lawrence Berkeley National Laboratory, Berkeley, California 94720} 
\affiliation{University of Liverpool, Liverpool L69 7ZE, United Kingdom} 
\affiliation{University College London, London WC1E 6BT, United Kingdom} 
\affiliation{Centro de Investigaciones Energeticas Medioambientales y Tecnologicas, E-28040 Madrid, Spain} 
\affiliation{Massachusetts Institute of Technology, Cambridge, Massachusetts  02139} 
\affiliation{Institute of Particle Physics: McGill University, Montr\'{e}al, Qu\'{e}bec, Canada H3A~2T8; Simon Fraser University, Burnaby, British Columbia, Canada V5A~1S6; University of Toronto, Toronto, Ontario, Canada M5S~1A7; and TRIUMF, Vancouver, British Columbia, Canada V6T~2A3} 
\affiliation{University of Michigan, Ann Arbor, Michigan 48109} 
\affiliation{Michigan State University, East Lansing, Michigan  48824}
\affiliation{Institution for Theoretical and Experimental Physics, ITEP, Moscow 117259, Russia} 
\affiliation{University of New Mexico, Albuquerque, New Mexico 87131} 
\affiliation{Northwestern University, Evanston, Illinois  60208} 
\affiliation{The Ohio State University, Columbus, Ohio  43210} 
\affiliation{Okayama University, Okayama 700-8530, Japan} 
\affiliation{Osaka City University, Osaka 588, Japan} 
\affiliation{University of Oxford, Oxford OX1 3RH, United Kingdom} 
\affiliation{Istituto Nazionale di Fisica Nucleare, Sezione di Padova-Trento, $^{ee}$University of Padova, I-35131 Padova, Italy} 
\affiliation{LPNHE, Universite Pierre et Marie Curie/IN2P3-CNRS, UMR7585, Paris, F-75252 France} 
\affiliation{University of Pennsylvania, Philadelphia, Pennsylvania 19104}
\affiliation{Istituto Nazionale di Fisica Nucleare Pisa, $^{ff}$University of Pisa, $^{gg}$University of Siena and $^{hh}$Scuola Normale Superiore, I-56127 Pisa, Italy} 
\affiliation{University of Pittsburgh, Pittsburgh, Pennsylvania 15260} 
\affiliation{Purdue University, West Lafayette, Indiana 47907} 
\affiliation{University of Rochester, Rochester, New York 14627} 
\affiliation{The Rockefeller University, New York, New York 10021} 
\affiliation{Istituto Nazionale di Fisica Nucleare, Sezione di Roma 1, $^{ii}$Sapienza Universit\`{a} di Roma, I-00185 Roma, Italy} 

\affiliation{Rutgers University, Piscataway, New Jersey 08855} 
\affiliation{Texas A\&M University, College Station, Texas 77843} 
\affiliation{Istituto Nazionale di Fisica Nucleare Trieste/Udine, I-34100 Trieste, $^{jj}$University of Trieste/Udine, I-33100 Udine, Italy} 
\affiliation{University of Tsukuba, Tsukuba, Ibaraki 305, Japan} 
\affiliation{Tufts University, Medford, Massachusetts 02155} 
\affiliation{Waseda University, Tokyo 169, Japan} 
\affiliation{Wayne State University, Detroit, Michigan  48201} 
\affiliation{University of Wisconsin, Madison, Wisconsin 53706} 
\affiliation{Yale University, New Haven, Connecticut 06520} 
\author{T.~Aaltonen}
\affiliation{Division of High Energy Physics, Department of Physics, University of Helsinki and Helsinki Institute of Physics, FIN-00014, Helsinki, Finland}
\author{J.~Adelman}
\affiliation{Enrico Fermi Institute, University of Chicago, Chicago, Illinois 60637}
\author{B.~\'{A}lvarez~Gonz\'{a}lez$^w$}
\affiliation{Instituto de Fisica de Cantabria, CSIC-University of Cantabria, 39005 Santander, Spain}
\author{S.~Amerio$^{ee}$}
\affiliation{Istituto Nazionale di Fisica Nucleare, Sezione di Padova-Trento, $^{ee}$University of Padova, I-35131 Padova, Italy} 

\author{D.~Amidei}
\affiliation{University of Michigan, Ann Arbor, Michigan 48109}
\author{A.~Anastassov}
\affiliation{Northwestern University, Evanston, Illinois  60208}
\author{A.~Annovi}
\affiliation{Laboratori Nazionali di Frascati, Istituto Nazionale di Fisica Nucleare, I-00044 Frascati, Italy}
\author{J.~Antos}
\affiliation{Comenius University, 842 48 Bratislava, Slovakia; Institute of Experimental Physics, 040 01 Kosice, Slovakia}
\author{G.~Apollinari}
\affiliation{Fermi National Accelerator Laboratory, Batavia, Illinois 60510}
\author{J.~Appel}
\affiliation{Fermi National Accelerator Laboratory, Batavia, Illinois 60510}
% ...

\collaboration{CDF Collaboration\footnote{With visitors from $^a$University of Massachusetts Amherst, Amherst, Massachusetts 01003,
$^b$Universiteit Antwerpen, B-2610 Antwerp, Belgium, 
$^c$University of Bristol, Bristol BS8 1TL, United Kingdom,
$^d$Chinese Academy of Sciences, Beijing 100864, China, 
$^e$Istituto Nazionale di Fisica Nucleare, Sezione di Cagliari, 09042 Monserrato (Cagliari), Italy,
$^f$University of California Irvine, Irvine, CA  92697, 
$^g$University of California Santa Cruz, Santa Cruz, CA  95064, 
$^h$Cornell University, Ithaca, NY  14853, 
$^i$University of Cyprus, Nicosia CY-1678, Cyprus, 
$^j$University College Dublin, Dublin 4, Ireland,
$^k$University of Edinburgh, Edinburgh EH9 3JZ, United Kingdom, 
$^l$University of Fukui, Fukui City, Fukui Prefecture, Japan 910-0017,
$^m$Kinki University, Higashi-Osaka City, Japan 577-8502,
$^n$Universidad Iberoamericana, Mexico D.F., Mexico,
$^o$University of Iowa, Iowa City, IA  52242,
$^p$Kansas State University, Manhattan, KS 66506,
$^q$Queen Mary, University of London, London, E1 4NS, England,
$^r$University of Manchester, Manchester M13 9PL, England,
$^s$Muons, Inc., Batavia, IL 60510, 
$^t$Nagasaki Institute of Applied Science, Nagasaki, Japan, 
$^u$University of Notre Dame, Notre Dame, IN 46556,
$^v$Obninsk State University, Obninsk, Russia,
$^w$University de Oviedo, E-33007 Oviedo, Spain, 
$^x$Texas Tech University, Lubbock, TX  79609, 
$^y$IFIC(CSIC-Universitat de Valencia), 56071 Valencia, Spain,
$^z$Universidad Tecnica Federico Santa Maria, 110v Valparaiso, Chile,
$^{aa}$University of Virginia, Charlottesville, VA  22906,
$^{bb}$Bergische Universit\"at Wuppertal, 42097 Wuppertal, Germany,
$^{cc}$Yarmouk University, Irbid 211-63, Jordan,
$^{kk}$On leave from J.~Stefan Institute, Ljubljana, Slovenia, 
}}
\noaffiliation

%% file: Paper_Final_body_mar10.tex
\begin{abstract}
We present a measurement of the top quark mass 
and of the top-antitop ($t\bar{t}$) pair production cross section
using $p\bar p$ data
collected with the CDF\,II detector at the 
Tevatron Collider at the Fermi National Accelerator Laboratory
and corresponding to an integrated luminosity
of 2.9 fb$^{-1}$.
We select events with six or more jets satisfying a number of kinematical requirements
imposed by means of
a neural-network algorithm. At least one of these jets must 
originate from a $b$ quark, as identified by the reconstruction 
of  a secondary vertex inside the jet. 
The mass measurement is based on a likelihood fit incorporating 
reconstructed mass distributions representative of signal and background, where
the absolute jet energy scale ($\mathrm{JES}$) is measured simultaneously with the top quark mass. 
The measurement yields a value 
 of $174.8\pm 2.4(\textnormal{stat+JES})~ ^{+1.2}_{-1.0}(\textnormal{syst})$\,GeV/$c^2$,
where the uncertainty from the absolute jet energy scale  is evaluated together 
with the statistical uncertainty. The procedure also measures the amount of signal
from which we derive a cross section, 
$\sigma_{t\bar{t}} = 7.2 \pm 0.5 (\textnormal{stat}) \pm 1.0 (\textnormal{syst}) \pm 0.4 (\textnormal{lum})$\,pb, 
for the measured values of top quark mass and JES.

 \end{abstract}

\maketitle

\section{Introduction}
\label{sec:Intro}


Since its early measurements, the large value of
the top quark mass ($M_{\mathrm{top}}$) has represented a really striking property of this particle, giving
to the top quark a special position within the standard model (SM) 
and suggesting also possible links to new physics\,\cite{sm_ext}.
In fact, apart from being itself a fundamental parameter of the SM, $M_{\mathrm{top}}$ is
by far the largest mass among the ones of the observed fermions, and this
makes the top quark contribution  dominant in higher order
corrections to many observables.
Therefore $M_{\mathrm{top}}$ plays a central role in checking the
consistency of theoretical predictions of the SM.
The higher order corrections apply also to the $W$ boson propagator, and therefore
affect the calculated value of the $W$ mass, $M_{W}$.
As the latter depends logarithmically on the mass of the 
Higgs boson, precise measurements of $M_{W}$ and $M_{\mathrm{top}}$ allow setting indirect
constraints on the value of the mass of this fundamental, but still unobserved
particle\,\cite{higgs}.
Moreover, possible contributions due to some unknown physics might also be constrained.
Finally, the present value of $M_{\mathrm{top}}$ makes the Yukawa coupling to the Higgs field of
${\cal O}(1)$  and this could indicate a special role of the top quark in
the mechanism of electroweak symmetry breaking.

All these reasons make the accurate knowledge of
$M_{\mathrm{top}}$ a really important issue, but the same is true for
the measurement of the $t\bar{t}$ production cross section ($\sigma_{t\bar{t}}$),
both as a test for physics contributions beyond the SM and as a test
of current next-to-leading-order (NLO) QCD calculations~\cite{theo-xsec}.
Usually, measurements of $\sigma_{t\bar{t}}$ rely upon event counting and
are performed assuming an {\it a priori} value for $M_{\mathrm{top}}$. The technique used here
allows the simultaneous measurement of
both these important and related properties of the top quark.

At the Tevatron Collider at Fermi National Accelerator Laboratory, top quarks are produced mostly in pairs.
In the SM  the top quark decays into a $W$ boson and a $b$ quark almost 100\% of the time,
and the topology of the final state resulting from 
a $t\bar{t}$ event depends on the hadronic or leptonic decay of the two 
final-state $W$ bosons.
In this paper, we consider events characterized by a multijet topology  ({\em all-hadronic} mode) with
no energetic leptons. This $t\bar t$ final state has the advantage of a large branching ratio 
($\approx 4/9$) and of 
having no undetectable final-state particles.
The major challenge of this channel is  the 
large background from  QCD multijet production, which dominates the signal by 3 orders 
of magnitude after the application of a specific online event selection (trigger). 
To increase the purity of the candidate sample, requirements based on the kinematical and topological 
characteristics of SM $\ttbar$ events are expressed in terms of an
artificial  neural network and applied to the data.
Further improvement is then obtained from the requirement of at least one jet identified as originating from 
a $b$ quark using a secondary vertex $b$-tagging algorithm.
Simulations predict that a clear $t\bar{t}$ signal 
will thus become visible over background in the selected
data sample, and the measurement of the  top quark mass and the $t\bar{t}$ 
cross section is made possible in spite of the overwhelming 
QCD multijet production.

A reconstructed top quark mass is determined by fitting the 
kinematics of the six leading jets in the event to a $t\bar{t}$ final state.
This variable, denoted as $m_{t}^{\mathrm{rec}}$, does not strictly represent a 
measurement of $M_{\mathrm{top}}$, but its distribution obtained 
by a sample of $t\bar{t}$ events is sensitive to $M_{\mathrm{top}}$ itself.  
The jet energy scale ($\mathrm{JES}$) is a factor representing
the set of corrections needed to obtain a better estimate of the energy of a parton 
starting from a jet reconstructed by clusters in the calorimeter. 
The default $\mathrm{JES}$ used in simulated events is obtained
by a tuning to the data, but possible discrepancies  between data and simulation lead 
to an uncertainty on this value. The strong correlation existing between  
the $m_{t}^{\mathrm{rec}}$ distribution and the $\mathrm{JES}$ implies therefore a corresponding 
uncertainty on $M_{\mathrm{top}}$.
However, the $\mathrm{JES}$ can be calibrated
using the selected samples of $t\bar{t}$ candidate events, where
a second variable, $m_{W}^{\mathrm{rec}}$, is reconstructed by the four-momenta of  
the jets assigned to the $W$ bosons. This variable is related to the well-known value 
of the $W$-boson mass, and the $\mathrm{JES}$ 
can be adjusted in such a way that both the $m_{t}^{\mathrm{rec}}$ and the $m_{W}^{\mathrm{rec}}$
distributions for simulated events match the observed data.
The inclusion of this procedure, usually referred to as {\em in situ} 
calibration,  enables a significant reduction of the systematic uncertainty 
associated with the inaccurate knowledge of the
$\mathrm{JES}$, and represents an important improvement of the work described 
in this paper with respect to the previous CDF analysis by a 
similar method\,\cite{ahprd}.
   
The $m_{t}^{\mathrm{rec}}$ and $m_{W}^{\mathrm{rec}}$ distributions
are reconstructed in two separate samples of selected data events, defined
by the presence of exactly one and two or more $b$-tagged jets respectively.
The data are then compared to corresponding distributions expected from background 
and  $t\bar t$ events simulated with various values of the top quark mass 
and of the $\mathrm{JES}$ to fit for these parameters.
In addition, the fitted signal yields are used to derive 
a measurement of the $t\bar{t}$ production cross section. 

The results reported here are based on data taken between March 2002 and April 2008, corresponding 
to an integrated luminosity of $2.9$~fb$^{-1}$.
This measurement complements  other recent determinations of the top quark mass and $t\bar{t}$ 
cross section by 
CDF and D0\,\cite{latestCDFD0mass,latestCDFD0xsec} in other final states, and improves  
the latest CDF measurements in the same channel\,\cite{ahprd, latestCDFah}.

The organization of the paper is as follows: Section\,\ref{sec:detector} contains a brief description
of the CDF\,II detector. The trigger and the neural-network-based sample selection are discussed in
Sec.\,\ref{sec:dataset}, along with the identification of jets initiated by $b$ quarks ($b$ jets).
Sections\,\ref{sec:MonteCarlo} and\,\ref{sec:bkg}  present the simulated signal samples
and
the data-driven method we use for estimating the background from multijet data. 
Section\,\ref{sec:temp} describes how the fundamental variables 
$m_{t}^{\mathrm{rec}}$ and $m_{W}^{\mathrm{rec}}$  are reconstructed, 
while in Sec.\,\ref{sec:cutoptimization} 
we present the final requirements to define the samples of events
used in the measurement.
The parametrization of the dependence of the distributions of reconstructed variables
on the values of the top quark mass and the jet energy scale are described in Sec.\,\ref{sec:temppar}.
The fit to the experimental distributions and its calibration are described
in Secs.\,\ref{sec:like} and\,\ref{sec:sanity}, respectively.
Section\,\ref{sec:syst} details the study of 
the systematic uncertainties on the mass measurement, that is then reported
in Sec.\,\ref{sec:meas}.
We describe in Sec.\,\ref{sec:xsec} the measurement of the $t\bar{t}$ cross section.


\section{The CDF\,II Detector}
\label{sec:detector}

The CDF\,II detector\,\cite{CDFdetector} is an azimuthally and forward-backward symmetric apparatus designed 
to study $p\bar p$ collisions at the Tevatron. A cylindrical coordinate system is used 
where $\theta$ is the polar angle to the
proton beam direction at the event vertex, $\phi$ is the azimuthal angle about the beam axis, and
pseudorapidity is defined as \mbox{$\eta = - \ln \left[ \tan(\theta/2)\right] $}.
We define transverse energy as $\Et = E \sin\theta$ and transverse momentum as $p_T = p\sin\theta$ 
where $E$ is the energy measured by calorimeters, and $p$ is the magnitude of the momentum measured 
by a tracking system.
The detector consists of a magnetic spectrometer surrounded by calorimeters and muon chambers. 
The charged particle tracking system is immersed in a 1.4~T solenoidal magnetic field with axis 
parallel to the beamline. A set of silicon microstrip detectors provides charged particle tracking in
the radial range from 1.5 to 28~cm, while 
a 3.1~m long open-cell drift chamber, the central outer tracker (COT), covers the radial range
from 40 to 137 cm. 
In combination the silicon and COT detectors provide excellent tracking up to about 
pseudorapidities $|\eta|\le 1.1$, and with decreasing precision up to $|\eta| \le 2.0$.
Segmented electromagnetic and hadronic calorimeters 
surround the tracking system, and measure the energy deposit of particles interacting in the calorimeters. 
The electromagnetic and hadronic calorimeters 
are lead-scintillator and iron-scintillator 
sampling devices, respectively, covering the range $|\eta|\le 3.6$. 
They are segmented in the central region ($|\eta|<1.1$) 
in towers of 15$^\circ$ in azimuth and 0.1 
in $\eta$, and the forward region ($1.1<|\eta|<3.6$) in towers of 7.5$^\circ$ for
 $|\eta|<2.11$ and 15$^\circ$ for $|\eta|>2.11$, 
while the coverage in $|\eta|$ increases 
gradually from 0.1 to 0.6\,.   
The electromagnetic calorimeters\,\cite{ecal,pem} are instrumented
with proportional chambers (at large angles) or
scintillating strip detectors (in the forward regions), which measure
the transverse profile of electromagnetic showers
at a depth corresponding to the expected shower maxima.
Drift chambers located outside the central hadronic calorimeters and behind a 60~cm iron shield 
detect muons with $|\eta| \le 0.6$\,\cite{CMU}. Additional drift chambers and scintillation  
counters detect muons in the region $0.6<|\eta|<1.5$. Multicell gas Cherenkov counters\,\cite{CLC} with a 
coverage of $3.7<|\eta|<4.7$ measure the average number of inelastic $p\bar p$ collisions and thereby 
are used to determine the luminosity.

 
\section{\boldmath Multijet Event Selection and $b$ tagging}
\label{sec:dataset}

The final state of all-hadronic $\ttbar$ events is characterized by the presence of at least 
six jets from the decay of the two top quarks, where additional jets might come from initial- or 
final-state radiation (ISR or FSR). Events having such a topology  are collected using a  
multijet trigger 
which relies on calorimeter information.
Subsequently, jets are identified during event reconstruction
 by grouping clusters of energy in the calorimeter using a  
fixed-cone algorithm with a  radius of 0.4 in  $\eta$-$\phi$ space\,\cite{jets}. 
After a preliminary selection of multijet events, a neural-network selection 
based on relevant  kinematical variables is used to further improve the purity of the sample. 


\subsection{Multijet Trigger}

The CDF trigger system has three levels.
The first two levels consist of 
special-purpose electronic circuits and the third one of conventional programmable digital processors.
At level 1, the trigger requires the presence of
at least one calorimeter tower with transverse energy
$E_{T}^{\mathrm{tow}}\ge 10$~GeV. At level 2 the total
transverse energy, obtained as the sum over all calorimeter
towers, $\sum E_{T}^{\mathrm{tow}}$, must be $\ge 175$~GeV. 
Moreover, the presence of  least four clusters of towers, 
each with transverse energy $E_{T}^{\mathrm{clus}}\ge~15$~GeV, is
required.
Finally, the third trigger level confirms the level 2 selection using a more accurate 
determination of the jet energy, requiring four or more reconstructed jets with 
$\Et\ge 10$~GeV. 
Approximately $14\,\times 10^{6}$ events satisfy the trigger requirements,
corresponding to an events signal-over-background ratio (S/B) of about 1/1200, assuming 
a theoretical cross section of $6.7$~pb for a top quark mass of 175 GeV/$c^2$\,\cite{theo-xsec}.


\subsection{Preselection and Topology  Requirements}
\label{sec:presel}

Events satisfying  the trigger requirements are reconstructed in terms of their final-state 
observables (tracks, vertices, charged leptons, and jets). We retain only those events that are 
well contained in detector acceptance, requiring the primary event vertex\,\cite{vertex} 
to lie inside the luminous region ($|z|<60$~cm). 
We remove 
events having well identified energetic electrons or muons as defined in\,\cite{ljetsKIN},
namely electrons with $E_{T} > 20$\,GeV and muons with  $p_{T} > 20$\,GeV/$c$.

In order to have jets matching as accurately as possible to the hard scattering partons, 
we correct jet energies  for detector response and multiple interactions\,\cite{JESNIM}. 
First, we consider the $\eta$-dependence of  detector response 
and energy loss in the uninstrumented regions.
Then, after accounting for the small extra energy deposited by multiple 
collisions in the same
beam-beam bunch crossing, a correction for calorimeter nonlinearity is applied so that 
the jet energies are equal,  on average, to the energy of the particles incident 
on  the jet cone.
The total uncertainty on the estimate of the original parton energy, 
where all uncertainties 
for the individual corrections are added in quadrature,
varies from 8\% to 3\% with jet transverse energy increasing from 15\,GeV to 50 GeV, 
and remains approximately 
constant at 3\% above 50 GeV. Jets with $|\eta|\le 2$ and $\Et\ge 15$\,GeV,
after all corrections are applied, are selected for further analysis.

As the uncertainty on the missing transverse energy, $\met$\,\cite{met}, 
increases proportionally to $\sqrt{\sum\Et}$\,\cite{metjetPRL}, 
its significance is defined as $\frac{\met}{\sqrt{\sum\Et}}$, 
where the $\met$ is corrected for
any identified muons,
while $\sum\Et$ is obtained by summing the $E_T$'s of all the selected jets.
We then require that $\frac{\met}{\sqrt{\sum\Et}}$ be $<3$\,${\rm GeV}^{\frac{1}{2}}$ 
to select events with small $\met$. 
At this stage, called preselection, 
we are left with about $8.2\, \times \,10^{6}$ events.

As the topology of the candidate events is determined by the jet multiplicity, 
we define the signal region by selecting events with a number 
of jets $6\le N_{\rm jets}\le 8$ and 
we also require jet pairs to be
separated by at least 0.5 units in the $\eta$-$\phi$ space.
The number of events passing these 
additional requirements is $1.671\, \times \,10^{6}$
with an expected S/B of approximately 1/430.


\subsection{Neural-Network-based Kinematical Selection}
\label{sec:sel}

To further improve the purity of the signal sample,
we use a multivariate
approach and take advantage of the distinctive features of signal and 
background events through a neural network, which takes into account the 
correlations between the kinematical 
variables which enter as input nodes in the network. 
The network uses the {\sc mlpfit} package\,\cite{mlp} as implemented by {\sc root}\,\cite{root} through 
the {\em TMultiLayerPerceptron} class. 
\par
A first set of 11 global variables, summarized in Table\,\ref{tab:nnvar}, have already been proven to be 
effective\,\cite{ahprd} in reducing the QCD background. 
Studies performed for this analysis on the jet development in the calorimeter 
have indicated that a good 
discrimination between quark-initiated and gluon-initiated jets can be accomplished  
with $\eta$ moments ($M_{\eta}$) and $\phi$ moments ($M_{\phi}$) of a jet,
which are defined as
%
\begin{equation}
M_\eta = \sqrt{ \left[\sum_{\mathrm{tow}} \frac{E_{T}^{\mathrm{tow}}}{E_T}\eta^2_{\mathrm{tow}} \right] - \eta^2 }
\end{equation}
%
and
%
\begin{equation}
M_\phi = \sqrt{\left[\sum_{\mathrm{tow}}  \frac{E_{T}^{\mathrm{tow}}}{E_T}\phi^2_{\mathrm{tow}} \right] - \phi^2 }\,,
\end{equation}
%
where $E_T$, $\eta$, and  $\phi$ are, respectively, the transverse energy, the pseudorapidity and
the azimuthal angle of the jet, while $E_{T}^{\mathrm{tow}}$ is the transverse energy deposited
in the calorimeter towers belonging to the jet.  

 We remove possible biases coming from $E_T$ distributions, which might differ in signal 
and background events, by deconvoluting the $E_T$ 
dependence  through a rescaling of all moments to a common reference value of $E_T=50$\,GeV. 
We obtain what we call scaled moments\,:
%
\begin{equation}
M^s_\eta = M_\eta \times \frac{f^\eta_q(50~ {\rm GeV})}{f^\eta_q(E_T)}
\end{equation}
%
and
%
\begin{equation}
M^s_\phi = M_\phi \times \frac{f^\phi_q(50~ {\rm GeV})}{f^\phi_q(E_T)}\,,
\end{equation}
where $f^\eta_q(E_T)$ and $f^\phi_q(E_T)$ are the functions that fit the profiles 
of $M_\eta$ $vs$ $E_T$ and of 
$M_\phi$ $vs$ $E_T$ in quark-initiated jets from simulated $t\bar t$  events.

\begin{figure}[htbp]
\centering

\includegraphics[width=8.5cm]{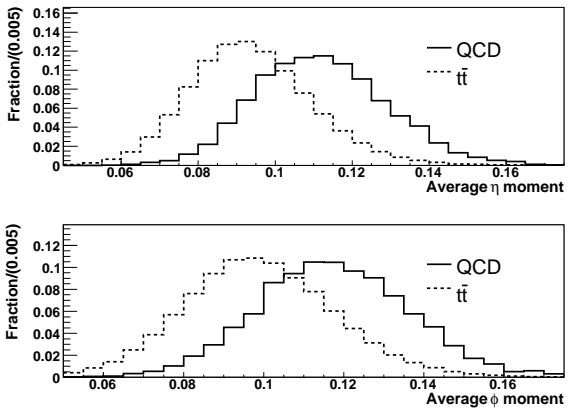}  

\caption{
         Geometric average of the $\eta$ scaled moments ($\langle M^{s}_{\eta} \rangle$, upper plot) and of the 
         $\phi$ scaled moments ($\langle M^{s}_{\phi} \rangle$, lower plot) 
         for QCD multijet (solid histogram) and 
         simulated $\ttbar$ (dashed histogram) events with $6\le N_{\mathrm{jets}}\le 8$. 
        } 

\label{fig:shapes}
\end{figure}

 These  scaled moments are quite different for jets coming from a quark or a gluon 
in simulated $t\bar t$ events. Such a behavior has been verified in data events 
where the jet origin is well known.
To take advantage of the large number of jets in a  $t\bar t$ event, we consider the geometric average 
of the $\eta$ moments and of  the $\phi$ moments, see Fig.\,\ref{fig:shapes}, evaluated using all jets 
which are not identified as coming from a heavy quark by the criteria 
explained in Sec.\,\ref{sec:btag}.

\begin{figure}[htbp]
\centering

\includegraphics[width=8.0cm]{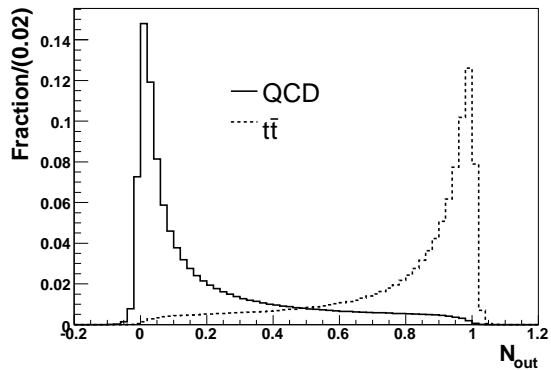}

\caption{
         Neural-network output $N_{\mathrm{out}}$ for QCD multijet (solid histogram) and 
         simulated $\ttbar$ (dashed histogram) events with $6\le N_{\mathrm{jets}}\le 8$. 
         Histograms are normalized to unity. The neural-network implementation 
         that we use in the 
         {\em TMultiLayer-Perceptron} produces
         an output which is not strictly bound between 0 and 1.
        }
 
\label{fig:nnout}
\end{figure}

The 13 variables are used as inputs to a neural network with two hidden layers with 20 and 10 hidden nodes, 
respectively, and one output node.
The network is trained on same-size  samples of signal and background events
 with $6\le N_{\rm jets}\le 8$  (about half a million events).
In order to model the signal we use the  {\sc pythia}\,v6.2\,\cite{Pythia}
 leading-order (LO) Monte Carlo generator with parton showering followed by a simulation of 
the CDF\, II detector. The reference top quark mass chosen for the training is 
$M_{\mathrm{top}}=175$\,GeV/$c^2$. The background is obtained from the multijet data 
events themselves, since the signal fraction is expected to be very small before applying
the neural-network selection. 
 The value of the output node, $N_{\mathrm{out}}$, is the quantity we use as a discriminator between
 signal and background, and is shown in  Fig.\,\ref{fig:nnout} for the $6\le N_{\mathrm{jets}}\le 8$ sample.

\begin{table}[hbtp]
\begin{center}
\caption{Input variables to the neural network.}\label{tab:nnvar}
\begin{tabular}{cl}
\hline\hline
\parbox[c][5mm][c]{15mm}{Variable} & ~~~~~~~~~~~~~~~~~~~~Description \\
\hline
\parbox[c][5mm][c]{15mm}{ $ \sum \Et    $ }             & Scalar sum of selected jets $E_T$\\
\parbox[c][5mm][c]{15mm}{ $ \sum _3\Et  $ }             & As above, except the two highest-$E_T$ jets\\
\parbox[c][5mm][c]{15mm}{ $ C           $ }             & Centrality\\
\parbox[c][5mm][c]{15mm}{ $ A           $ }             & Aplanarity\\
\parbox[c][5mm][c]{15mm}{ $ M_{2j}^{\mathrm{min}}$ }    & Minimum dijet invariant mass\\
\parbox[c][5mm][c]{15mm}{ $ M_{2j}^{\mathrm{max}}$ }    & Maximum dijet invariant mass\\
\parbox[c][5mm][c]{15mm}{ $ M_{3j}^{\mathrm{min}}$ }    & Minimum trijet invariant mass\\
\parbox[c][5mm][c]{15mm}{ $ M_{3j}^{\mathrm{max}}$ }    & Maximum trijet invariant mass\\
\parbox[c][5mm][c]{15mm}{ $ E_T^{\star, 1}$ }           & $E_T\sin^2\theta^\star$ for the highest-$E_T$ jet\\
\parbox[c][5mm][c]{15mm}{ $ E_T^{\star, 2}$ }           & $E_T\sin^2\theta^\star$ for the next-to-highest-$E_T$ jet\\
\parbox[c][5mm][c]{15mm}{ $ \langle E_T^\star\rangle$ } &  Geometric mean over the remaining jets\\
\parbox[c][5mm][c]{15mm}{ $ \langle M_{\eta}^s\rangle$} & Geometric mean over the untagged jets\\
\parbox[c][5mm][c]{15mm}{ $ \langle M_{\phi}^s\rangle$} & Geometric mean over the untagged jets\\
\hline\hline
\end{tabular}
\end{center}
\end{table}

\subsection{\boldmath Tagging $b$ quarks}
\label{sec:btag}

In order to enrich the $t\bar t$ content in the event sample, we use  a $b$-tagging algorithm based on 
secondary vertex reconstruction 
as described in detail in\,\cite{vertex,massTMT}. The algorithm identifies
a jet  likely to contain a hadron with a $b$ quark
by reconstructing 
its decay vertex with at least two
high-quality tracks with hits in the silicon vertex  detector. 
A $b$-tagged jet ({\em tag}, in brief) must have
an associated secondary vertex with a displacement from the primary vertex 
in the transverse
plane larger than 7.5 times the transverse-displacement resolution. 
This is evaluated for each secondary vertex, but its typical
value is about 190\,$\mu$m.  
The tagging efficiencies for jets coming from the fragmentation of  $b$ or $c$ quarks
are corrected in simulated events according to the efficiency seen in  the data, 
by a factor $0.95\pm 0.04$, both for $b$ jets
and $c$ jets. These factors are  described in detail in\,\cite{vertex}.


\section{Event Simulation}
\label{sec:MonteCarlo}

The standard model $t\bar{t}$ events used to study the event selection
and to check the performance of the method (Sec.\,\ref{sec:sanity}) are simulated using
{\sc pythia}\,v6.2\,\cite{Pythia}. Samples generated with input values of the top quark mass, 
$M_{\mathrm{top}}^{\mathrm{in}}$, ranging from 160 to 190 GeV$/c^{2}$ are considered and, for each sample, the event selection
is repeated by varying the $\mathrm{JES}$ from its default value\,\cite{JESNIM}.
The displacement, denoted as $\Delta\mathrm{JES}$, is measured relative to 
the uncertainty, $\sigma_{\mathrm{JES}}$, on the default value itself, so that
the value of $\mathrm{JES}$ applied to simulated events is increased by
$\Delta\mathrm{JES} \cdot  \sigma_{\mathrm{JES}} $ with respect to  the default. 
To test the method, input values $\Delta\mathrm{JES}^{\mathrm{in}}$ ranging from $-3$ to $+3$ 
are considered.  

Different generators and different values for the model parameters are 
used to estimate the systematic uncertainties, as described in Sec.\,\ref{sec:syst}.



\section{Background Estimate}
\label{sec:bkg}

The background for the  $\ttbar$ multijet final state comes  mainly from QCD 
production of 
heavy-flavor quark pairs ($b\bar b$ and $c\bar c$) and 
from false tags of light-flavor quark jets. 
Other standard model processes such as $W/Z$+jets have  a smaller
production cross section and small acceptance due to the selection requirements. 

Given the large theoretical uncertainties on the QCD multijet production cross section,
a more accurate background estimate is 
obtained from the data, rather than from  Monte Carlo simulations. 
A tag rate per jet, defined as the probability of tagging a jet whose tracks 
are reconstructed in the vertex detector ({\it fiducial} jet), is then evaluated in 
a sample of events with exactly four jets passing the preselection
and therefore still dominated by the background (S/B  $\approx 1/5000$).
The rate is parametrized in terms 
of variables sensitive to both the tagging efficiency for heavy-flavored objects 
and the probability of false tags\,: the jet $\Et$, the number of tracks reconstructed in the 
silicon vertex detector  and associated with the jet, $N_{\rm trk}^{\mathrm{jet}}$, 
and the number of primary vertices reconstructed in
the event, $N_{\rm vert}$~\cite{ahprd}. 
By definition, the tag rate estimates the probability that a fiducial jet
having, on average, the characteristics of jets from background events is tagged.   
Its average value is about $3.7\%$, with negligible uncertainty. 
%
%
However, direct exploitation of the tag rate to predict the number of background events 
with exactly a given number of tags would give incorrect numbers.
This happens because, by construction, this rate is the ratio between the
number of tagged jets and the number of fiducial jets in a whole sample of events.
Possible correlations among jets in the same event are not considered. 
As  heavy-flavor quarks come in pairs in QCD background, the probability 
to tag a pair of jets 
in the same event is therefore larger than the simple product of the tag probabilities 
of individual jets given by the tag rate. 

To account for this we introduce correction factors to obtain
a better estimate for the number of 1-tag and $\ge 2$-tag background events.
These factors are derived in a control sample dominated by the background
(events with six, seven or eight jets and $N_{\mathrm{out}} \le 0.25$, with S/B\,$\approx 1/1300$ for 
one tag and S/B$\approx 1/400$ for $\ge 2$ tags)
as the ratio between the observed number of events with $n$ tags (with $n = 1,\,2,\,3$)
and the average expectation obtained
by using the tag rate to evaluate the probability for each event to have 
the same number, $n$, of tagged jets.
These factors represent, therefore, average corrections 
to the probability for a possible {\em tag configuration},
that is for the assumption that among the fiducial jets in an event of the sample
selected before the $b$-tagging requirements ({\em pretag} sample) 
only a given subset is actually tagged when the algorithm is applied. 
Their average values are $0.94$, $1.48$, and $2.46$ for events with one, two, and three tagged 
jets, with relative statistical uncertainties of $0.4\%$, $1.1\%$, and  $5.1\%$ respectively.
Similarly to the tag rate, these corrections should be valid for events with the
characteristics of background events. 

The accuracy of our modeling of the background processes is verified in control samples,
i.e. on events with  
higher values of $N_{\mathrm{out}}$ and therefore with a larger
fraction of signal events and with possible different kinematics and background composition.
As the background prediction is performed using the data in the pretag sample,
the presence of $t\bar{t}$ events must also be taken into account.
Therefore a correction is applied to derive a better evaluation, $n_{(b,\,\mathrm{exp})}$, of the background
normalization from the raw estimate $n_{(b,\,\mathrm{raw})}$ directly
obtained by the corrected tag rate matrix.
This correction must subtract the contribution, $n_{t\bar{t}}^{tr} $, 
coming from applying the matrix to signal events and included in $n_{(b,\,\mathrm{raw})}$. 
Denoting by $N_{\mathrm{obs}}$ the number of events observed in the data sample, 
by $n_{t\bar{t}}$ the number of signal events in this sample and
assuming that the excess of events with respect to the expected background 
is totally due to the signal, the correction can be written as\,:
%
\begin{eqnarray}
  n_{(b,\,\mathrm{exp})} & = &  n_{(b,\,\mathrm{raw})} - n_{t\bar{t}}^{tr} \nonumber \\
               & = &  n_{(b,\,\mathrm{raw})} - \frac{ n_{t\bar{t}}^{tr} }{ n_{t\bar{t}} } \cdot n_{t\bar{t}}  \nonumber \\
               & = &  n_{(b,\,\mathrm{raw})} - \frac{ n_{t\bar{t}}^{tr} }{ n_{t\bar{t}} } \cdot (N_{\mathrm{obs}} - n_{(b,\,\mathrm{exp})} ) \,,~~~~~
\end{eqnarray}
%
%
which, with $R_{t\bar{t}} \equiv  n_{t\bar{t}}^{tr} /   n_{t\bar{t}}  $, gives
%
\begin{eqnarray}
    n_{(b,\,\mathrm{exp})}  &  =  &  \frac{  n_{(b,\,\mathrm{raw})} - R_{t\bar{t}} \cdot  N_{\mathrm{obs}}  }{ 1 - R_{t\bar{t}}  } \,. 
\end{eqnarray}
%
$R_{t\bar{t}}$ can be inferred from simulated events and amounts to
$0.314 \pm 0.003$\,(stat) [$0.067 \pm 0.0014$\,(stat)] for 1-tag ($\geq 2$-tag) events. 
Further possible discrepancies between the observed and expected number of
events are  considered as due to the modeling of the background
and accounted for as a systematic uncertainty.


\section{Mass reconstruction}
\label{sec:temp}

The simultaneous measurement of the top quark mass and the $\mathrm{JES}$ 
is based on the reconstruction, event by event, 
of both the top quark and the $W$ masses through a constrained fitting technique.
The shapes of the distributions obtained by this procedure are sensitive to
the values of both $M_{\mathrm{top}}$ and $\mathrm{JES}$. Therefore, for simulated events,
they are built using samples corresponding to the different input values of $M_{\mathrm{top}}$ and $\Delta\mathrm{JES}$
listed in Sec.\,\ref{sec:MonteCarlo}.

Moreover, given the different resolution in the reconstructed top quark mass and the $W$ boson mass, 
and also the different S/B which can be achieved by requiring events with exactly  one or  $\ge 2$ tags,
two sets of distributions are separately derived in these samples.

\subsection{Reconstructed top quark mass}
\label{subsec:Mtoptemplates}

For each event we determine a reconstructed top quark mass, $m_t^{\mathrm{rec}}$, 
from the four-momenta of selected jets.
Sixteen equations can be considered to connect the four-momenta of the two top quarks and their
decay products according to the 
\mbox{$t \bar t \rightarrow b \bar b \, W^+ W^- \rightarrow b \bar b \, q_1 \bar q_2 \, q_3 \bar q_4$} hypothesis\,:
\begin{eqnarray}
 p_t^\mu        & = &  p_{W^+}^\mu + p_b^\mu \,,        \\
 p_{\bar t}^\mu & = &  p_{W^-}^\mu + p_{\bar b}^\mu\,,   \\
 p_{W^+}^\mu    & = &  p_{q_1}^\mu + p_{\bar q_2}^\mu \,, \\
 p_{W^-}^\mu    & = &  p_{q_3}^\mu + p_{\bar q_4}^\mu \,, 
\end{eqnarray} 
%
with $\mu   =  0,\,1,\,2,\,3$. 
There are 13 unknown quantities, i.e., the unknown top quark mass and the three-momenta of 
the top quarks and of the $W$ bosons, 
so the kinematics of the events are overconstrained.

The fit is performed using only the six highest-$E_T$ jets (leading jets) of the 
event and considering their possible
assignments to quarks of a $t \bar t$ final state. The total number of different permutations
giving two doublets of jets corresponding to the $W$ bosons and two triplets of jets corresponding
to the top quarks is 90. Since we require the presence of $b$ tags, assigning
the tagged jets only to $b$ quarks reduces this number to 30 for 1-tag events and 6 
in case of two or more $b$ tags\,\cite{twotagfit}. 

For each permutation 
the kinematics of the event is reconstructed minimizing the following $\chi^2$ function\,:
\begin{eqnarray}
\chi^{2} &  =  &  ~\frac{ \big(m_{jj}^{(1)}-M_W \big)^2 }{ \Gamma^2_W } +
                  \frac{ \big(m_{jj}^{(2)}-M_W \big)^2 }{ \Gamma^2_W }  \nonumber \\
         &  +   &  \frac{ \big(m_{jjb}^{(1)}-m_{t}^{\mathrm{rec}} \big)^2 }{ \Gamma^2_t } +
                   \frac{ \big(m_{jjb}^{(2)}-m_{t}^{\mathrm{rec}} \big)^2}{\Gamma^2_t}~~~~~~~   \nonumber \\
         &  +   &  \sum^6_{i=1} \frac{ \big(p^{\mathrm{fit}}_{T,i}-p^{\mathrm{meas}}_{T,i} \big)^2}{\sigma^2_i} \,. 
\end{eqnarray}
The minimization procedure is performed with respect to seven parameters, i.e., the reconstructed top quark mass 
$m_{t}^{\mathrm{rec}}$ and the transverse momenta $p^{\mathrm{fit}}_{T,i}$ of
the six jets, which  are constrained 
to the measured value $p^{\mathrm{meas}}_{T,i}$ within their known resolution $\sigma_{i}$.
The invariant masses of the jet doublets assigned to light-flavor quarks coming from a $W$, $m_{jj}^{(1,2)}$,
and of the trijet systems including one doublet and one of the jets assigned to $b$ quarks, $m_{jjb}^{(1,2)}$,
are evaluated by the trial momenta of jets at each step of the minimization.
On the contrary, the measured mass $M_W$  and the natural width $\Gamma_W$ of the $W$ boson
as well as the assumed natural width of the 
top quark, $\Gamma_t$, are kept constant to $80.4\,$GeV/$c^2$, $2.1$ GeV/$c^2$ and $1.5$\,GeV/$c^2$ 
respectively\,\cite{pdg,widt}.

The permutation of jets which gives the lowest $\chi^2$ value is selected, and the
corresponding fitted value of  $m_{t}^{\mathrm{rec}}$ enters 
an invariant mass distribution ({\em template}) 
which will be used for the $M_{\mathrm{top}}$ measurement.


\subsection{\boldmath Reconstructed $W$ mass}
\label{subsec:Mwtemplates}

Reconstructing the mass of $W$ bosons by using dijet systems represents a possibility to 
obtain a variable,
in principle, insensitive to $M_{\mathrm{top}}$ which allows, therefore,
an independent determination of $\mathrm{JES}$.

To build the $m_{W}^{\mathrm{rec}}$ distributions we use the same procedure and $\chi^2$ expression  
considered for  $m_{t}^{\mathrm{rec}}$,
but now the $W$-boson mass is also left as a free parameter in the fit 
(i.e. $M_{W}$ becomes $m_{W}^{\mathrm{rec}}$).
Again, for each event, the value of $m_{W}^{\mathrm{rec}}$ corresponding to the permutation of the
jet-to-parton assignments with the lowest $\chi^2$ enters the distribution.

Using different fits in the reconstruction of  $m_t^{\mathrm{rec}}$ and $m_{W}^{\mathrm{rec}}$ can lead to selecting
different assignments of jets to partons for the two variables in the same event. 
This is not 
a problem as the same procedure is followed both on data and simulated events. 
Reconstructing the top quark
mass using a constant value of $M_{W}$, as decribed in Sec.\,\ref{subsec:Mtoptemplates}, improves the
resolution of the distributions and therefore the determination of the {\em true} value of $M_{\mathrm{top}}$.
The correlations between the values of $m_t^{\mathrm{rec}}$ and $m_{W}^{\mathrm{rec}}$ in the same event are taken into
account in the calibration of the likelihood fit used for the measurement (Sec.\,\ref{sec:like}).


\subsection{Background templates}
\label{sec:bkgtemplates}

In order to reconstruct data-driven background templates we apply the kinematical fitter
to the sample of events passing the neural-network selection, but before the requirement 
of tagged $b$ jets.

The same procedures described in Secs.~\ref{subsec:Mtoptemplates} 
and~\ref{subsec:Mwtemplates} are repeated
on these events assigning fiducial jets to $b$ quarks and then looping over all possible 
assignments of other jets to the remaining quarks, performing the fit
for each permutation and selecting the reconstructed $m_t^{\mathrm{rec}}$ and $m_W^{\mathrm{rec}}$ values 
corresponding to the best $\chi^{2}$. These values then enter the templates weighted 
by the {\em corrected} probability 
of the assumed tag configuration; see Sec.~\ref{sec:bkg}.
As for the normalization, the
background distributions also need to be corrected 
for the presence of signal in the pretag sample by subtracting
the contribution from $t\bar{t}$ events. 
The shape of this contribution is obtained from simulated samples
and depends on the assumed $M_{\mathrm{top}}$ and $\mathrm{JES}$, while the normalization
is given by the difference $n_{(b,\,\mathrm{raw})} - n_{(b,\,\mathrm{exp})}$, 
as described in Sec.\,\ref{sec:bkg}.

In order to check how well our modeling describes the background, we consider events in  
control regions defined by the $N_{\mathrm{out}}$ value, in ranges where the signal presence after 
tagging is still very low. 
In these regions the templates, i.e. the main elements of our measurement, are reconstructed
by the procedure described in the previous sections, both for the signal 
and the background, as well as other important distributions like $N_{\mathrm{out}}$ and
the $\chi^{2}$ of the fit used to build the $m_{t}^{\mathrm{rec}}$ templates.
These distributions are then compared to observed data, taking into account the
contribution from signal events. 
The agreement is generally good in all the control regions,
and this confirms the reliability of the background model.

\begin{figure}[htbp!]
 \begin{center}
   \begin{tabular}{c}
\epsfig{file=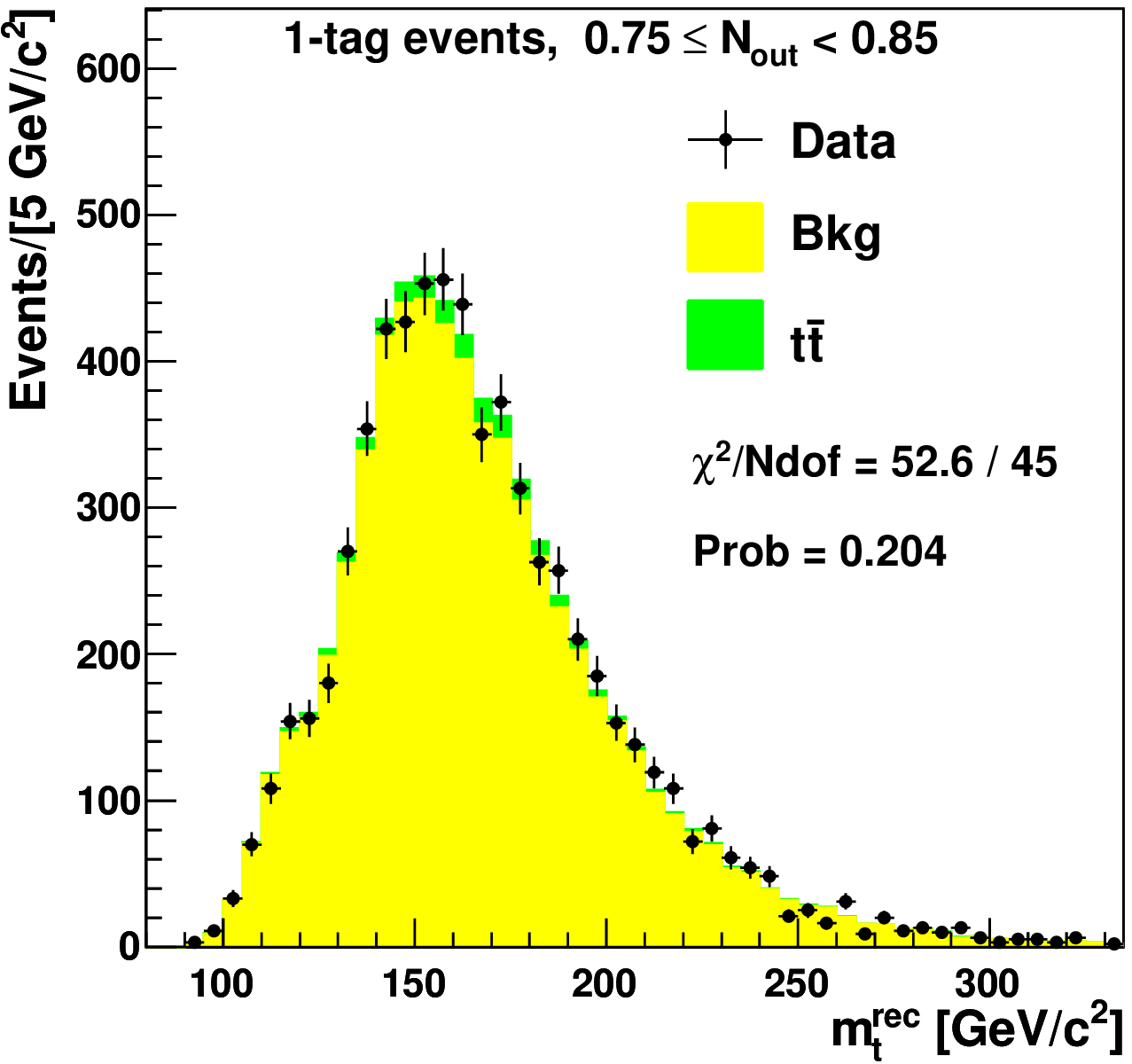,width=8cm}   \\
\epsfig{file=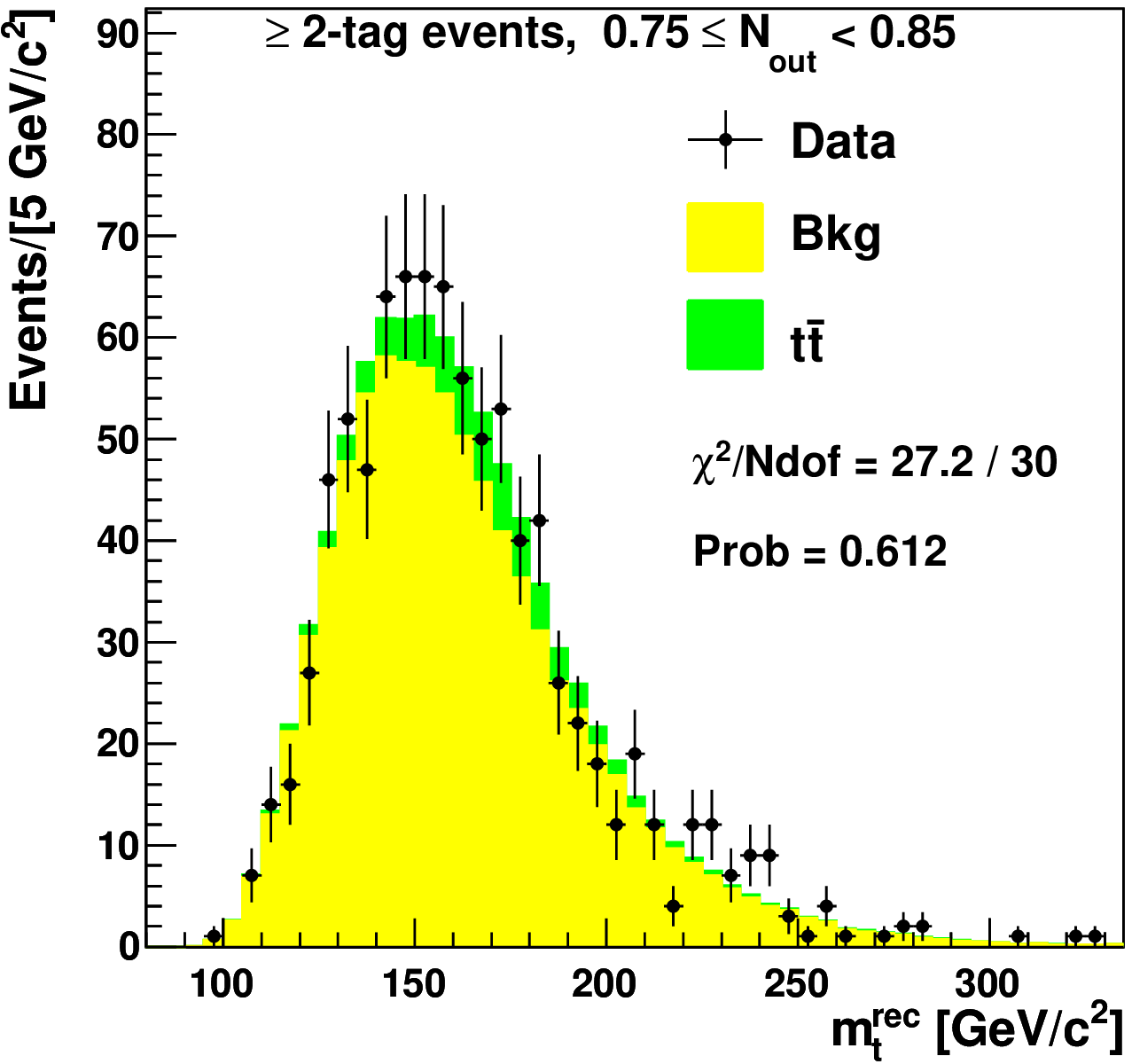,width=8cm}
   \end{tabular}
   \caption{
            Histograms of the reconstructed top quark mass $m_{t}^{\mathrm{rec}}$ for 1-tag events, upper plot,
            and $\ge 2$-tag events, lower plot, are shown in a control region defined by
            $0.75 \le N_{\mathrm{out}} < 0.85$.
            Along with the data are plotted the 
            expected background and the signal contribution
            for $M_{\mathrm{top}}=175$\,GeV/$c^{2}$ and the default value $\Delta\mathrm{JES} = 0$,
            normalized to the difference between the data and the background.
            The value of the purely statistical
            $\chi^{2}$ probability is reported on each plot.
           }
  \label{fig:MtopValid}
 \end{center}
\end{figure}

\begin{figure}[htbp!]
 \begin{center}
   \begin{tabular}{c}
   \epsfig{file=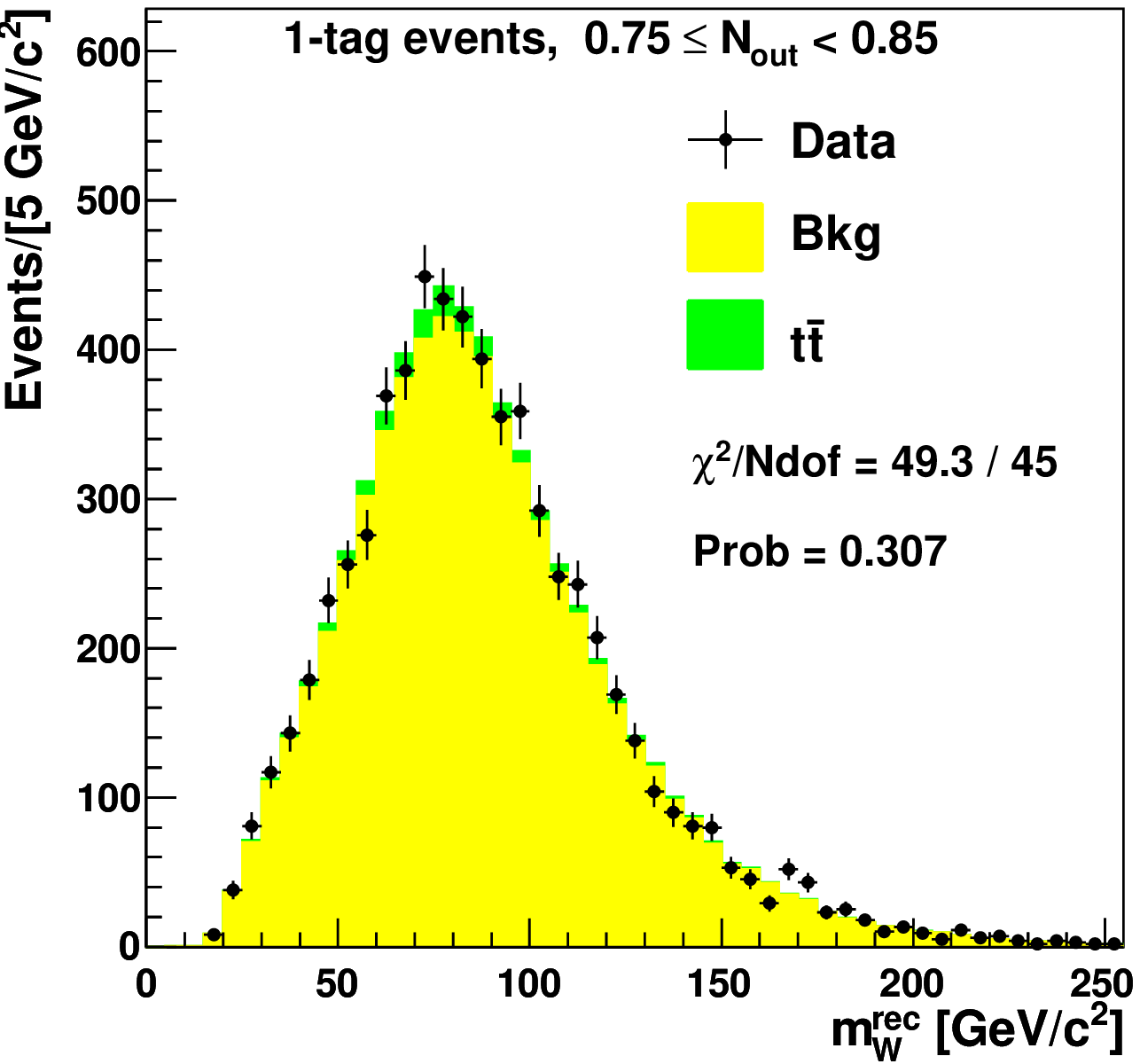,width=8cm}   \\
   \epsfig{file=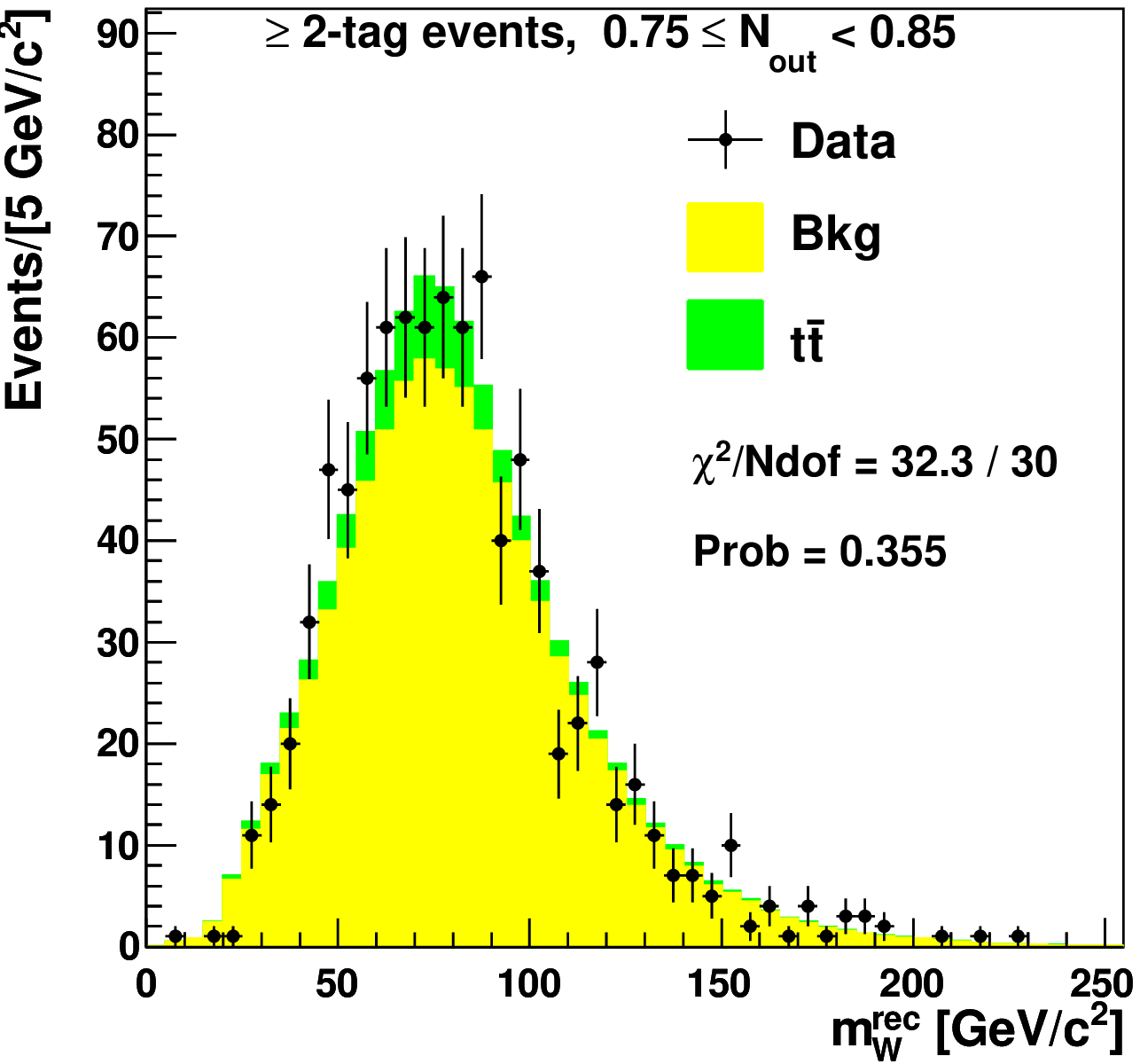,width=8cm}
   \end{tabular}
   \caption{
            Histograms of the reconstructed $W$ mass $m_{W}^{\mathrm{rec}}$ for 1-tag events, upper plot,
            and $\ge 2$-tag events, lower plot, are shown in a control region defined by
            $0.75 \le N_{\mathrm{out}} < 0.85$.
            Along with the data are plotted the 
	    expected background and the signal contribution
            for $M_{\mathrm{top}}=175$\,GeV/$c^{2}$ and the default value $\Delta\mathrm{JES} = 0$, 
            normalized to the difference between the data and the background.   
            The value of the purely statistical
            $\chi^{2}$ probability is reported on each plot.        
           }
  \label{fig:MwValid}
 \end{center}
\end{figure}

Figures~\ref{fig:MtopValid} and~\ref{fig:MwValid} show, as examples,
distributions of $m_{t}^{\mathrm{rec}}$ and $m_{W}^{\mathrm{rec}}$ 
in one of the control regions
for 1-tag and $\ge 2$-tag events, where the sum of signal and background
is compared to the same distributions reconstructed in the data. In these plots the 
integral of the signal distributions 
corresponding to  $M_{\mathrm{top}}=175$\,GeV/$c^{2}$ and the default
value $\Delta\mathrm{JES}$=0
have been normalized to the difference between the observed data and
the corrected expected background.


%

\section{Event samples}

\label{sec:cutoptimization}

In order to obtain the best performance from our method, we performed sets of {\em pseudoexperiments} (PEs)
to find out which requirements on the 
values of $N_{\mathrm{out}}$ and of the $\chi^{2}$ used to obtain
the $m_{t}^{\mathrm{rec}}$ values minimize the statistical uncertainty on the top quark mass measurement.
The procedure is 
similar to the one outlined in Sec.~\ref{sec:sanity},
with a binned version of the same likelihood. 
It is applied separately to 1-tag and $\geq 2$-tag samples and considers many different combinations
of possible requirements.
The smallest values for the uncertainty are obtained using  $\left(N_{\mathrm{out}} \ge 0.90,\,\chi^{2} \le 6\right)$
in the 1-tag sample and  $\left(N_{\mathrm{out}} \ge 0.88,\,\chi^{2} \le 5\right)$ in the $\ge 2$-tag sample so that
we add these requirements to the prerequisites described in Sec.~\ref{sec:presel}.
The final definition of the samples used in our analysis is summarized in Table~\ref{tab:samples}.

\begin{table}[htbp]
 \begin{center}
   \caption{Final definition and requirements for selected event samples.}
     \label{tab:samples} 
  \begin{tabular}{lccc}
    \hline \hline
 Event Sample~~~   & $b$-tags      & \parbox[c][5mm][c]{20mm}{$N_{\mathrm{out}}$}  &  $m_{t}^{\mathrm{rec}}$ fit $\chi^{2}$ \\
     \hline
 One tag          &  $\equiv 1$   & \parbox[c][5mm][c]{20mm}{     $ \ge 0.90 $    }   &   $ \le 6  $      \\
  $\ge 2$-tags  &  $ 2$ or $3$  & \parbox[c][5mm][c]{20mm}{     $ \ge 0.88 $    }   &   $ \le 5  $      \\ 
    \hline \hline
  \end{tabular}
 \end{center}
\end{table}

%
%
%
%
%
\noindent After these selections, 3452 and 441 events are observed for the 1-tag and $\ge 2$-tag samples 
respectively. We can evaluate the average number of background events expected in 
the selected samples and their uncertainties, as described in Sec.~\ref{sec:bkg}.
The systematic uncertainties on the background normalizations are estimated
by assuming that the discrepancy between the observed number of events in the data and
the sum of the expected contributions from signal and background 
(where, in this case, the theoretical cross section value of
 $6.7$\,pb is considered for $t\bar{t}$ events production) is due to a bad evaluation 
of the background.
This is done separately for 1-tag and $\ge 2$-tag samples, and the resulting 
relative uncertainties on the expected number of events are
\mbox{$\sigma( n_{(b,\,\mathrm{exp})}^{1\,tag} ) = 2.9\%$} 
and \mbox{$\sigma( n_{(b,\,\mathrm{exp})}^{\ge 2\,tags}) = 14.6\%$}, respectively.
The efficiencies of the full selection on $t\bar{t}$ events corresponding
to $M_{\mathrm{top}} = 175$\,GeV/$c^{2}$ and $\Delta\mathrm{JES} = 0$ are $3.6\%$ and $1.0\%$  for 1-tag 
and $\ge 2$-tag events respectively.
These values are used to evaluate the expected signal contributions of Table\,\ref{tab:finalsamples},
where $\sigma_{t\bar{t}} = 6.7$\,pb is assumed. 
In the same table, the observed number of events and the expected background in each sample are also summarized. 
\begin{table}[htbp]
 \begin{center}
   \caption{Number of events observed in the selected data samples and corresponding
            expected numbers of background and $t\bar{t}$ events. The signal contribution
            is evaluated for $M_{\mathrm{top}} = 175$\,GeV/$c^{2}$, $\Delta\mathrm{JES} = 0$ and $\sigma_{t\bar{t}} = 6.7$\,pb.
           }
     \label{tab:finalsamples} 
  \begin{tabular}{lccc}
    \hline \hline
                 Event sample   & \parbox[c][5mm][c]{20mm}{Observed}   &  Background  & \parbox[c][5mm][c]{17mm}{$t\bar{t}$}\\
     \hline
                 One tag          & \parbox[c][5mm][c]{20mm}{     $3452$    }   &   $2785 \pm 83$ &  $693$     \\
                 $\ge 2$ tags   & \parbox[c][5mm][c]{20mm}{     $ 441$    }   &   $ 201 \pm 29$ &  $193$    \\ 
    \hline \hline
  \end{tabular}
 \end{center}
\end{table}
%


\section{Likelihood fit}
\label{sec:likefit}

The technique described in Sec.\,\ref{sec:temp} 
allows one to obtain sets of  {\em observed} $m_t^{\mathrm{rec}}$ and $m_W^{\mathrm{rec}}$
values reconstructed in the data samples with 1 or $\ge 2$ tags
as well as to build signal and background distributions for the same
variables. 
In order to measure the top quark 
mass simultaneously with the $\mathrm{JES}$, a fit is performed  where an 
unbinned likelihood function is maximized
to find the  values of $M_{\mathrm{top}}$, $\Delta\mathrm{JES}$,\, and the number of 
signal ($n_{s}$) and background ($n_{b}$) 
events for each tagging category which give 
the probability density functions (p.d.f.'s)
best describing the data. 

\subsection{Probability density functions}
\label{sec:temppar}

\begin{figure}[tbp!]
\begin{center}
   \begin{tabular}{c}

    \epsfig{file=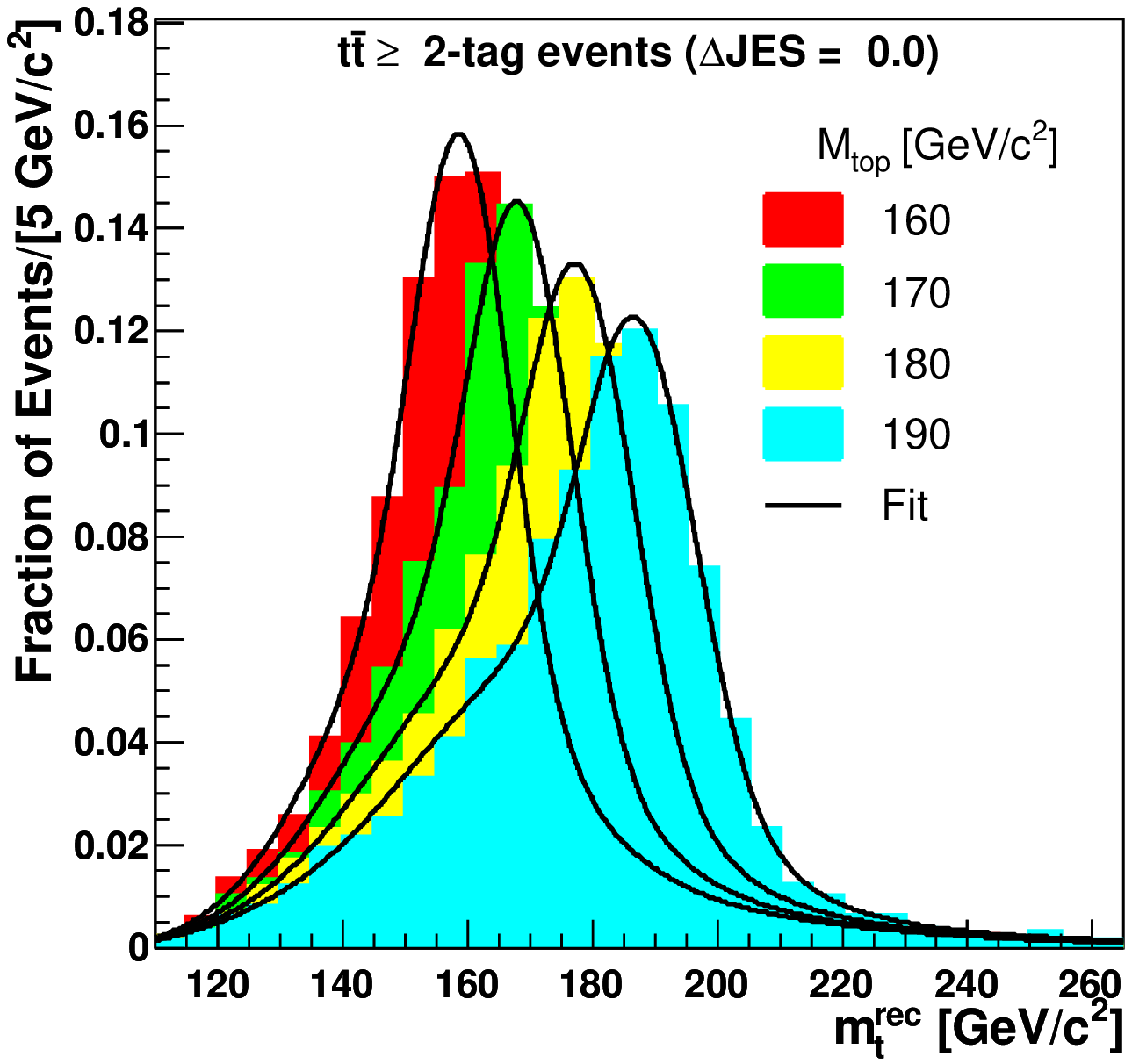,width=8cm}

   \\
  
    \epsfig{file=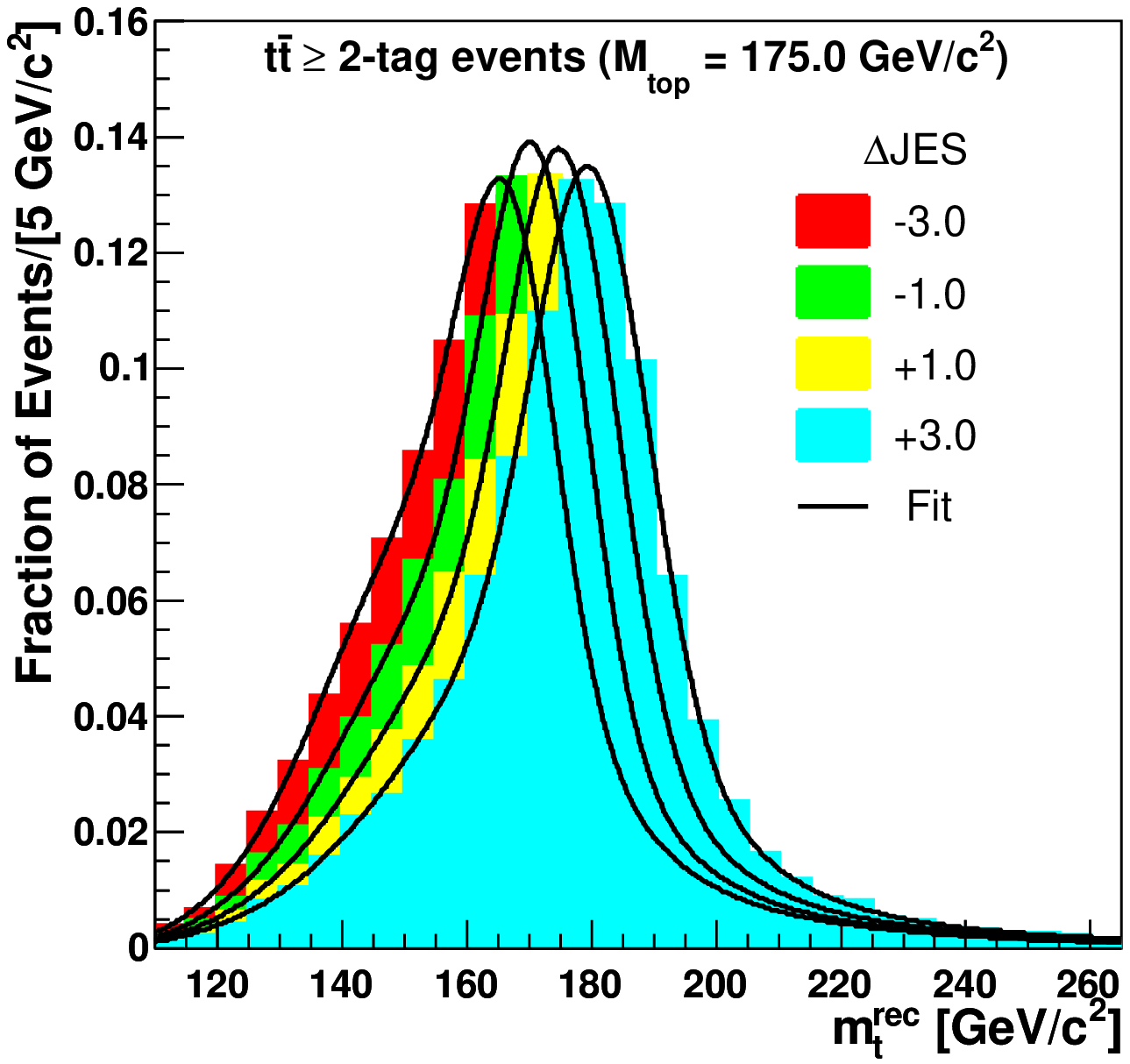,width=8cm}
    
   \end{tabular}
\caption{ 
         Histograms and corresponding fitted probability density functions for the 
          signal $m_{t}^{\mathrm{rec}}$ in 
         $\ge 2$-tag events  
         for a constant $\Delta\mathrm{JES}$ value ($\Delta\mathrm{JES}=0$), 
         varying the input top quark mass (upper plot)
         and for a constant $M_{\mathrm{top}}$ value (175\,GeV/$c^2$), varying the input jet energy scale 
         (lower plot).         
        } 
\label{fig:pdftop}
\end{center}
\end{figure}

\begin{figure}[htbp!]
\begin{center}
   \begin{tabular}{c}

    \epsfig{file=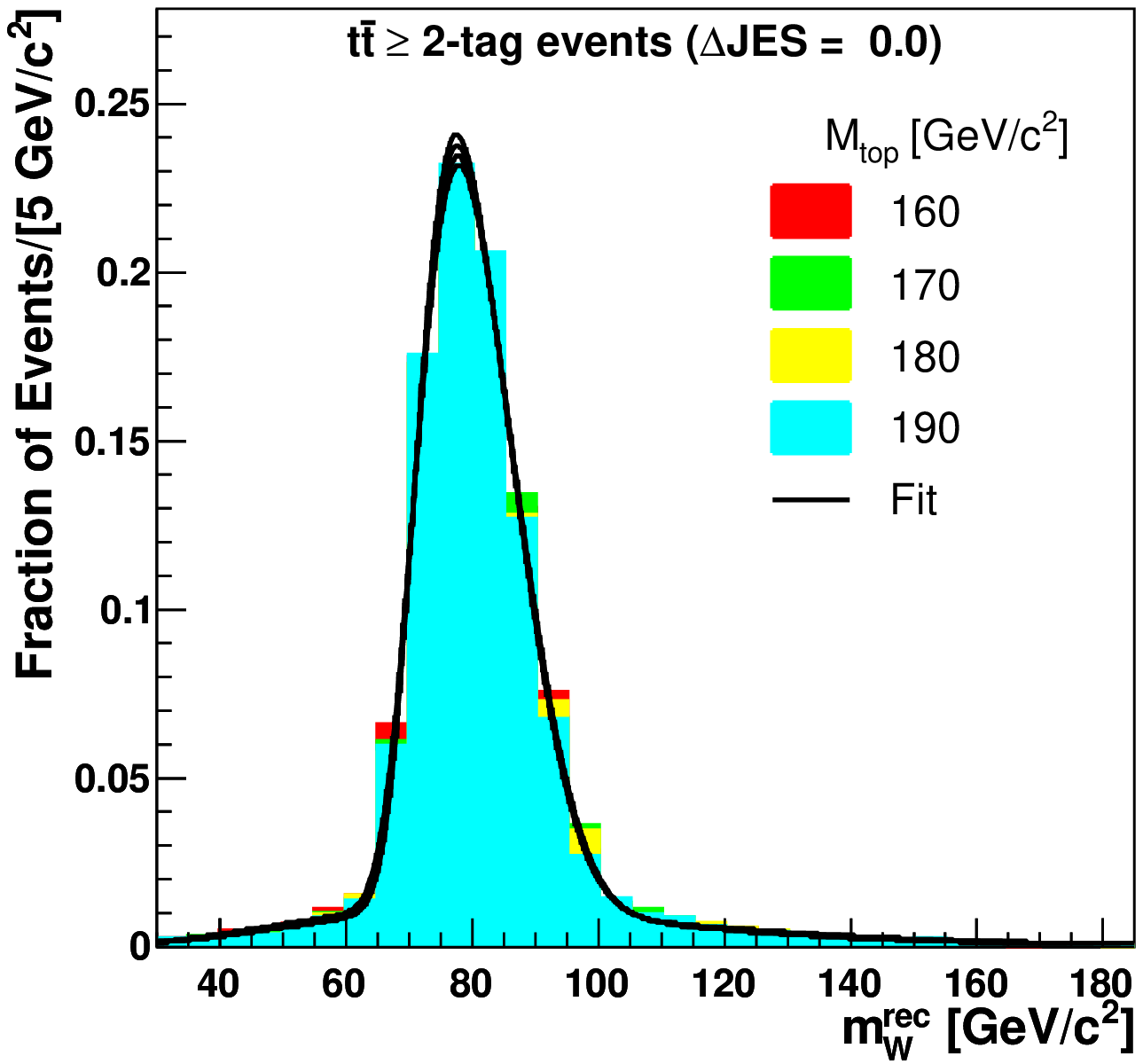,width=8cm}

   \\

    \epsfig{file=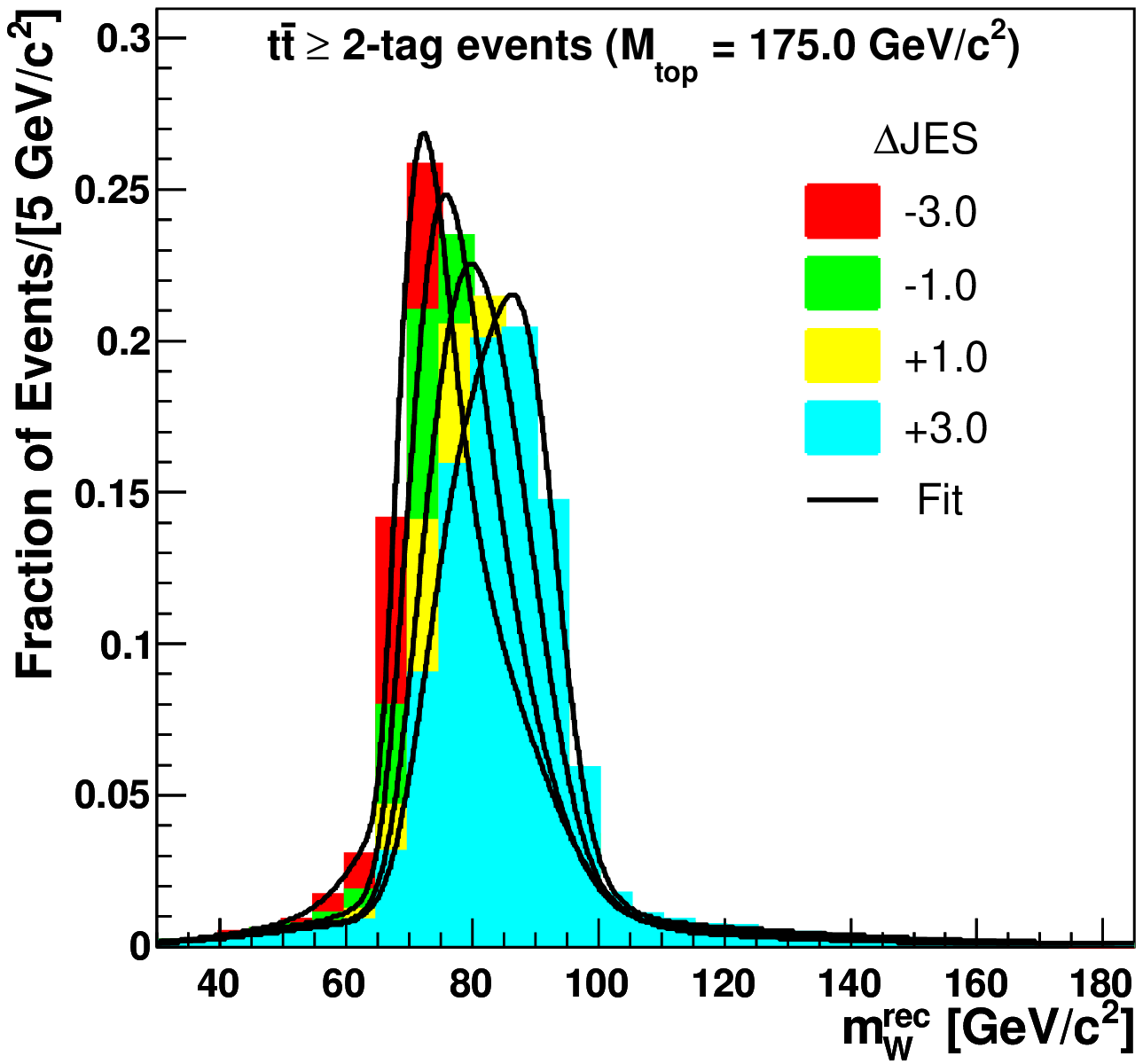,width=8cm}
    
   \end{tabular}
\caption{ 
         Histograms and corresponding fitted probability density functions 
         for the signal $m_{W}^{\mathrm{rec}}$ in 
         $\ge 2$-tag events  
         for a constant $\Delta\mathrm{JES}$ value ($\Delta\mathrm{JES}=0$), 
         varying the input top quark mass (upper plot),
         where the independence of $m_{W}^{\mathrm{rec}}$ on $M_{\mathrm{top}}$ is apparent,
         and for a constant $M_{\mathrm{top}}$ value (175\,GeV/$c^2$), varying the input jet energy scale 
         (lower plot).       
       } 
\label{fig:pdfw}
\end{center}
\end{figure}

The signal templates are fitted by normalized combinations
of Gamma and Gaussian p.d.f.'s, and the dependence of the shape 
on input $M_{\mathrm{top}}$ and $\Delta\mathrm{JES}$ is included writing 
the parameters of the  p.d.f.'s as linear functions of these
variables.
Figures~\ref{fig:pdftop} and~\ref{fig:pdfw} 
show examples of the fitted p.d.f.'\,s superimposed on the 
$m_{t}^{\mathrm{rec}}$ and $m_{W}^{\mathrm{rec}}$  
signal templates respectively for different $M_{\mathrm{top}}$ and $\Delta\mathrm{JES}$ values. 

\begin{figure}[th!]
 \begin{center}
    \begin{tabular}{c}

    \epsfig{file=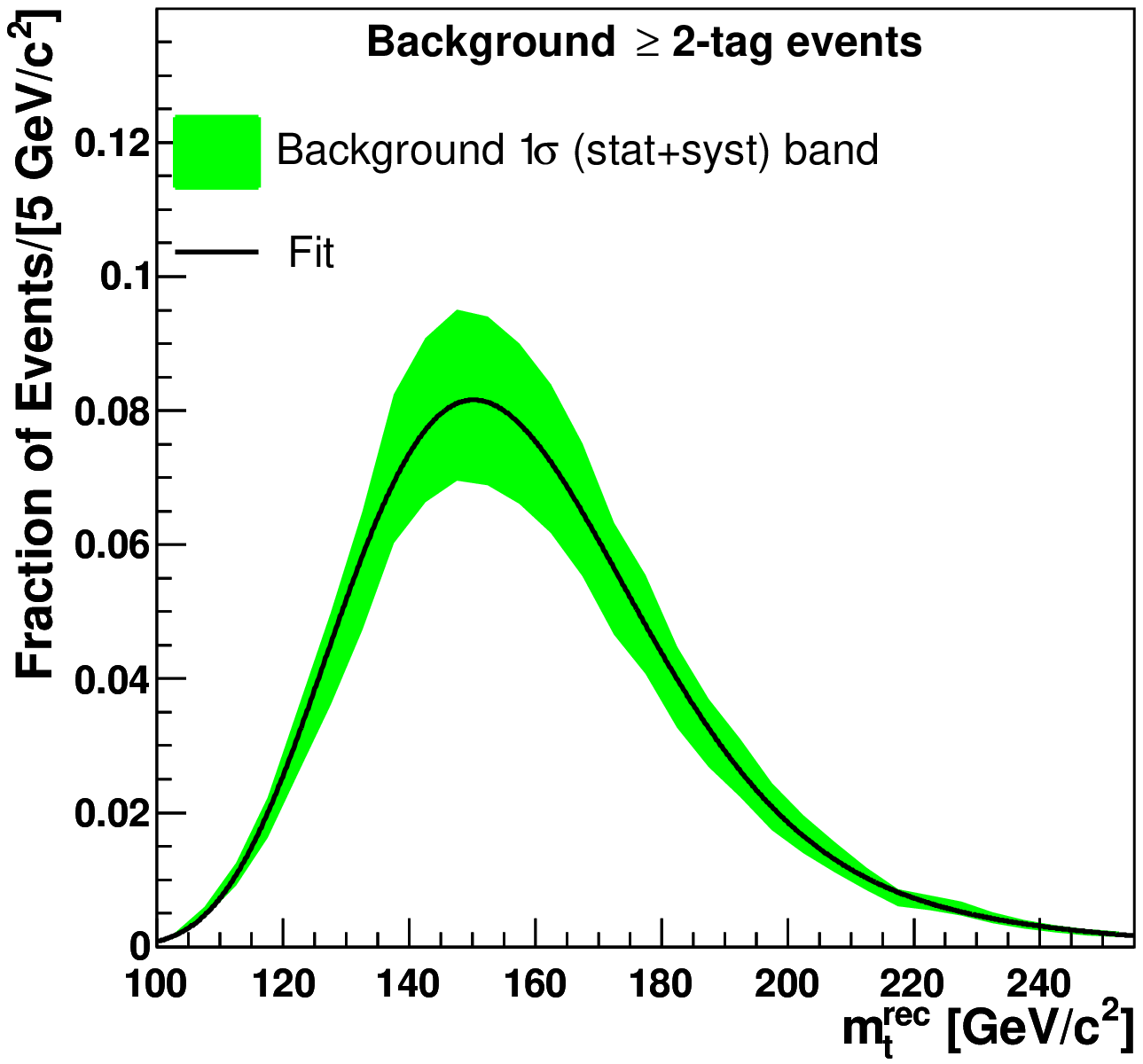,width=8cm} \\

    \epsfig{file=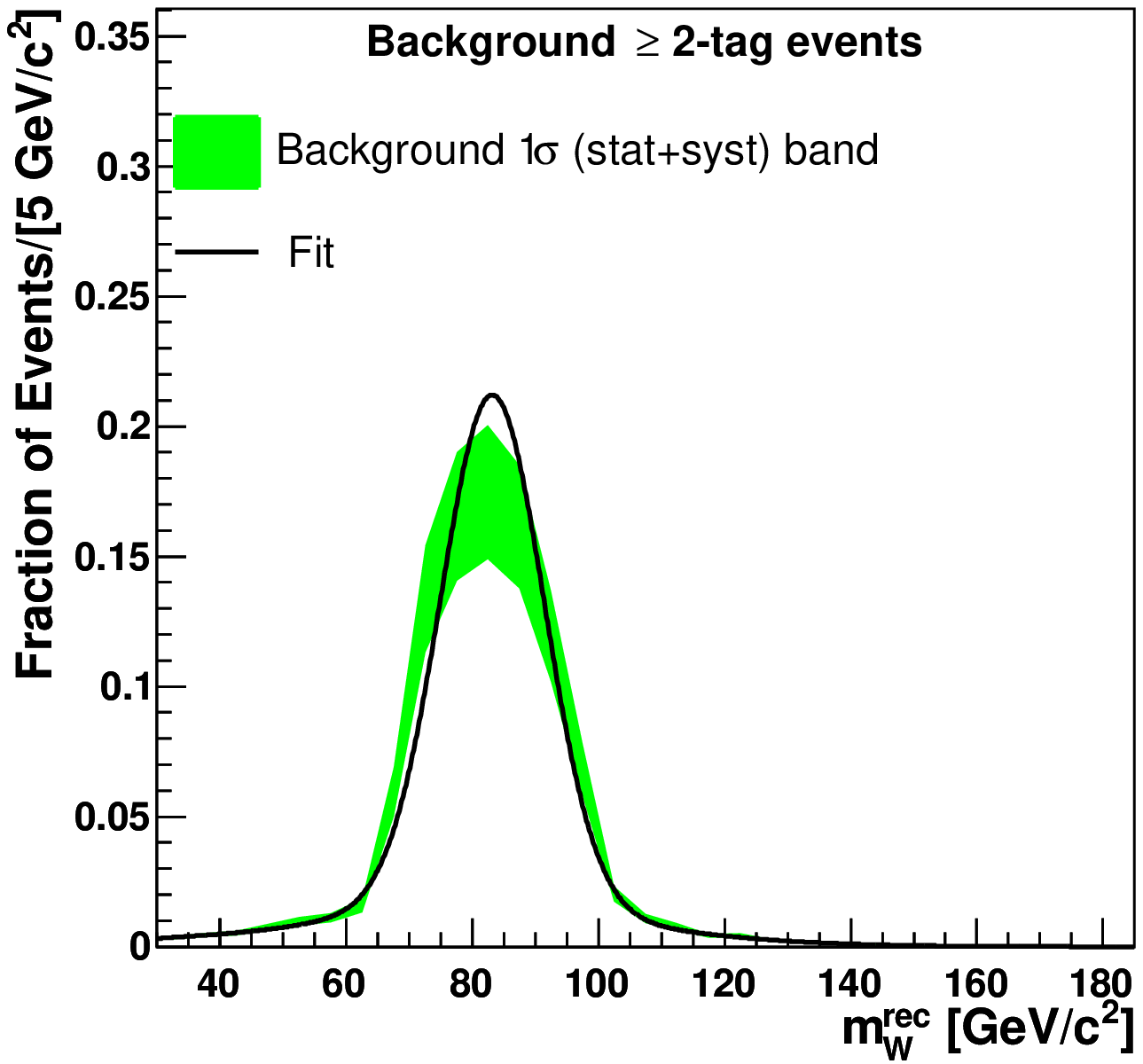,width=8cm}

   \end{tabular}
    \caption{  
              Data-driven background histograms of $m_{t}^{\mathrm{rec}}$ (upper plot) 
              and $m_{W}^{\mathrm{rec}}$ (lower plot) for $\ge 2$-tag events.
              The  bands denote the 1$\sigma$ uncertainty on the bin contents 
              of the histograms, including both statistical and
              systematic contributions. 
              The solid lines show the p.d.f.'s fitted to the histograms.
            }

\label{fig:pdfbkg}

\end{center}
\end{figure}

The shape of distributions built for the background cannot depend on the characteristics
of signal events, and in particular on the value of top quark mass. 
Moreover, as they are obtained from data, the shapes correspond to the 
reference value of the jet energy scale. For these reasons
no dependence on $M_{\mathrm{top}}$ and $\mathrm{JES}$ is considered in the p.d.f.'s used
to fit the background templates.  
Actually, a very weak dependence is introduced through the corrections to the shape 
of the background distributions, performed to take into account the presence of signal events
in the pretag sample, as described in Sec.\,\ref{sec:bkgtemplates}. 
These effects are taken into account as a systematic uncertainty.  
Examples of background $m_{t}^{\mathrm{rec}}$ and $m_{W}^{\mathrm{rec}}$ distributions and the 
corresponding fitted p.d.f.'s are shown 
in Fig.~\ref{fig:pdfbkg} for $\ge 2$-tag events. Discrepancies between the
fitted p.d.f.'s and the corresponding distributions are considered in the 
calibration procedure, presented in Sec.\,\ref{sec:sanity}.   

\subsection{The Likelihood Function}
\label{sec:like}

The likelihood function ${\cal L}$  is divided into three main parts and can be written as
\begin{eqnarray}
\label{eq:totallike}
{\cal L} = {\cal L}_{1\,\mathrm{tag}} \times {\cal L}_{\ge 2 \,\mathrm{tags}} \times {\cal L}_{\Delta\mathrm{JES}_{\mathrm{constr}}}~~~ \,.
\end{eqnarray}

The ${\cal L}_{1}$ and ${\cal L}_{\ge 2 \, \mathrm{tags}}$ terms further consist of other factors\,:  
\begin{eqnarray}
 {\cal L}_{1,\ge 2\, \mathrm{tags}} = {\cal L}_{M_{\mathrm{top}}} \times {\cal L}_{\mathrm{JES}} \times {\cal L}_{\mathrm{Poiss}}
    \times {\cal L}_{N^{\mathrm{bkg}}_{\mathrm{constr}}}\,,~~~  
\end{eqnarray}
where the four terms on the right side
assume respectively the following form [the superscripts referring to the
tag sample are omitted and $f_{s} \equiv n_{s} / (n_{s} + n_{b})$, $f_{b} \equiv 1 -  f_{s}$]\,:
{\footnotesize
\begin{eqnarray}
\label{eq:mtreclike}
%
\prod_{i=1}^{N_{\mathrm{obs}}} f_{s} \cdot P^{m_{t}^{\mathrm{rec}}}_{\mathrm{sig}}(m_{t,\,i}\,|\,M_{\mathrm{top}},\,\Delta\mathrm{JES})
                      + f_{b} \cdot P^{m_{t}^{\mathrm{rec}}}_{\mathrm{bkg}}(m_{t,\,i})\,, ~~~~~
\end{eqnarray}
}
{\footnotesize
\begin{eqnarray}
\label{eq:mwreclike}
\prod_{i=1}^{N_{\mathrm{obs}}} f_{s} \cdot P^{m_{W}^{\mathrm{rec}}}_{\mathrm{sig}}(m_{W,\,i}\,| \, M_{\mathrm{top}},\, \Delta\mathrm{JES})
                             + f_{b} \cdot P^{m_{W}^{\mathrm{rec}}}_{\mathrm{bkg}}(m_{W,\,i})\,, ~~~
\end{eqnarray} 
} 
{\small
\begin{eqnarray}
\label{eq:poisslike}
\frac{e^{-(n_s + n_b)} \cdot (n_s + n_b)^{N_{\mathrm{obs}}}}{N_{\mathrm{obs}}!} \,,
\end{eqnarray}
}  
{\small
\begin{eqnarray}
\label{eq:bkgconstlike}
\exp{\left[- ~ \frac{\left(n_{b}-n_{(b,\,\mathrm{exp})}\right)^2}{2\sigma_{n_{(b,\,\mathrm{exp})}}^2}\right]}\,. 
\end{eqnarray}
}
In expression~(\ref{eq:mtreclike}) the probability to observe the set 
$m_{t,\,i},~(i=1,...,N_{\mathrm{obs}})$ of $m_{t}^{\mathrm{rec}}$ values 
reconstructed in the data is calculated by using the total probability density 
function resulting from the combination
of the parametrized signal and background p.d.f.'s (Sec.\,\ref{sec:temppar}), 
$P^{m_{t}^{\mathrm{rec}}}_{\mathrm{sig}}$ and $P^{m_{t}^{\mathrm{rec}}}_{\mathrm{bkg}}$ respectively, 
as a function of the free parameters of the fit.    
In term~(\ref{eq:mwreclike}) the same is done for the set of the observed 
$W$ masses, $m_{W,\,i},~(i=1,...,N_{\mathrm{obs}})$,
and the $m_{W}^{\mathrm{rec}}$ p.d.f.
The term~(\ref{eq:poisslike}), ${\cal L}_{\mathrm{Poiss}}$, gives the probability to observe the
number of events selected in the data, given the average number of
signal ($n_{s}$) and background ($n_{b}$) events expected in the sample, as assumed at each step of the
likelihood fit.  In the last term,~(\ref{eq:bkgconstlike}),
the parameter $n_{b}$ is constrained by a Gaussian to the {\em a priori} background estimate given in
Sec.~\ref{sec:cutoptimization}, i.e. $n_{(b,\,\mathrm{exp})} = 2785\, \pm \, 83$ for 1-tag events and
$n_{(b,\,\mathrm{exp})} = 201\, \pm \, 29$ for $\ge 2$-tag events.    
Finally, the last term in expression~(\ref{eq:totallike}), ${\cal L}_{\Delta\mathrm{JES}_{\mathrm{constr}}}$, 
is a Gaussian term constraining
$\Delta\mathrm{JES}$ to its {\em a priori} value\,: 
\begin{eqnarray}
%
%
%
\exp{\left[- ~ \frac{\left( \Delta\mathrm{JES} - \Delta\mathrm{JES}_{\mathrm{constr}}\right)^{2}}{2}\right]}\,. 
\end{eqnarray}
When the measurement is performed on data, the $\mathrm{JES}$ can be constrained
to the value independently measured in~\cite{JESNIM}.
Given the meaning of $\Delta\mathrm{JES}$, described in Sec.\,\ref{sec:MonteCarlo},
this means that, in this case, $\Delta\mathrm{JES}_{\mathrm{constr}} = 0$.

%
%

\section{Verification and Calibration of the Method}
\label{sec:sanity}

We want to investigate the possible presence of biases in the top quark mass and 
jet energy scale measurements introduced 
by our method, as well as to have an estimate of its statistical power 
before performing the measurement on the actual data sample.
To do so, we run realistic PEs where {\em pseudodata} are extracted 
from simulated signal and data-driven background distributions.
A set of 3000 PEs is performed for each simulated value of the top quark mass and
of the displacement in the jet energy scale (Sec.\,\ref{sec:MonteCarlo}). 
Using the notation introduced in Sec.\,\ref{sec:MonteCarlo}, we refer to these input 
values as $M_{\mathrm{top}}^{\mathrm{in}}$ and $\Delta\mathrm{JES}^{\mathrm{in}}$ and they represent  the
{\em true}  values we want to measure.
In each PE the actual numbers of signal ($N_{(s,\,obs)}$) and 
background ($N_{(b,\,obs)}$) events in each 
tagging category are generated with Poisson distributions
with mean $n_{(s,\,\mathrm{exp})} =  N_{\mathrm{obs}} - n_{(b,\,\mathrm{exp})}$ and $n_{(b,\,\mathrm{exp})}$ respectively,   
where  $N_{\mathrm{obs}}$ are the observed number of events in the data samples 
($N_{\mathrm{obs}} = 3452 $ for 1-tag and 
 $N_{\mathrm{obs}} = 441 $ for $\ge 2$-tag events). 
A set of $N_{(s,\,obs)}$ and $N_{(b,\,obs)}$ mass values is then drawn
from $m_{t}^{\mathrm{rec}}$ and $m_{W}^{\mathrm{rec}}$ distributions of signal and background 
and used as input to the likelihood fit (Sec.\,\ref{sec:likefit}) that returns
simultaneous measurements of $M_{\mathrm{top}}$ and $\Delta\mathrm{JES}$, denoted as 
$M_{\mathrm{top}}^{\mathrm{out}}$ and $\Delta\mathrm{JES}^{\mathrm{out}}$.
The average of these measurements over the whole set of 3000 PEs represents the best
estimate of the input values obtained by the fitting procedure
and therefore can be used  to study its behavior.
We fit the dependence of these averages
with respect to the input values over the whole range of 
simulated $M_{\mathrm{top}}^{\mathrm{in}}$ and $\Delta\mathrm{JES}^{\mathrm{in}}$ as
{\small
\begin{eqnarray}
  \langle M_{\mathrm{top}}^{\mathrm{out}} \rangle~~  & = & (A_{00} + A_{01} \cdot \Delta\mathrm{JES}^{\mathrm{in}}) \nonumber\\
                   & + & (A_{10} + A_{11} \cdot \Delta\mathrm{JES}^{\mathrm{in}}) \cdot (M_{\mathrm{top}}^{\mathrm{in}}  - 175)
\,,~~~~~~~ \label{eq:MtopCalib} 
\end{eqnarray}
}
{\small
\begin{eqnarray}
 \langle \Delta\mathrm{JES}^{\mathrm{out}} \rangle  & = & \left[ B_{00} + B_{01} \cdot (M_{\mathrm{top}}^{\mathrm{in}}  - 175)\right] \nonumber \\
                           & +  & 
                           \left[ B_{10} + B_{11} \cdot (M_{\mathrm{top}}^{\mathrm{in}}  - 175)\right] \cdot \Delta\mathrm{JES}^{\mathrm{in}}\,.~~~~~~~ \label{eq:JESCalib}  
%
\end{eqnarray}
}

These relations can be inverted to obtain calibration functions to be applied to further measurements
and therefore, on average, a more reliable estimate of the true values (2D calibration).
The calibrated values resulting from  a measurement giving $M_{\mathrm{top}}^{\mathrm{out}}$ 
and $\Delta\mathrm{JES}^{\mathrm{out}}$
are denoted as $M_{\mathrm{top}}^{\mathrm{corr}}$ and $\Delta\mathrm{JES}^{\mathrm{corr}}$, 
while the respective uncertainties,
obtained by propagating through the calibration the uncertainties from the likelihood fit,
are $\delta M_{\mathrm{top}}^{\mathrm{corr}}$ and $\delta \Delta\mathrm{JES}^{\mathrm{corr}}$.
A second set of PEs is then performed to test the goodness of the procedure. 
Table~\ref{tab:calib} shows the coefficients $A_{ij}$ and $B_{ij}$ obtained both
from calibrated and uncalibrated PEs compared to their ideal values in the absence
of any bias.

\begin{table}[htbp]
 \begin{center}
   \caption{Coefficients of expressions~(\ref{eq:MtopCalib}) and~(\ref{eq:JESCalib})
            as obtained from calibrated and uncalibrated pseudoexperiments. 
            The ideal values  in the absence of any bias are also shown. 
           }
     \label{tab:calib} 
  \begin{tabular}{lccc}
    \hline \hline
\parbox[c][4.5mm][c]{10mm}{}   &  \parbox[c][4.5mm][c]{32mm}{Uncalibrated}  &  \parbox[c][4.5mm][c]{22mm}{Calibrated}  &  
                                                         \parbox[c][4.5mm][c]{12mm}{Ideal} \\
              &                               PEs             &     PEs      &    value          \\
    \hline
 $A_{00}$     &  \parbox[c][4.5mm][c]{30mm}{$175.47  \pm 0.01$}   &  $174.99 \pm 0.01 $  & $175$  \\
 $A_{01}$     &  \parbox[c][4.5mm][c]{30mm}{$-0.24   \pm 0.01$}   &  $  0.00 \pm 0.01 $  & $0$    \\
\hline
 $A_{10}$     &  \parbox[c][4.5mm][c]{30mm}{$0.985   \pm 0.002$}  &  $1.000  \pm 0.002$  & $1$    \\
 $A_{11}$     &  \parbox[c][4.5mm][c]{30mm}{$0.009   \pm 0.001$}  &  $0.000  \pm 0.001$  & $0$    \\
\hline
 $B_{00}$     &  \parbox[c][4.5mm][c]{30mm}{$-0.026  \pm 0.003$}  &  $0.002  \pm 0.003$   & $0$    \\
 $B_{01}$     &  \parbox[c][4.5mm][c]{30mm}{$0.0009  \pm 0.0004$} &  $-0.0001 \pm 0.0004$ & $0$    \\
\hline
 $B_{10}$     &  \parbox[c][4.5mm][c]{30mm}{$1.052   \pm 0.002$}  &  $0.999  \pm 0.002$   & $1$    \\
 $B_{11}$     &  \parbox[c][4.5mm][c]{30mm}{$-0.0016 \pm 0.0002$} &  $0.0001 \pm 0.0002$  & $0$    \\

    \hline \hline
  \end{tabular}
 \end{center}
\end{table}
 
\begin{figure}[th!]
\begin{center}
\begin{tabular}{c}

   \epsfig{file=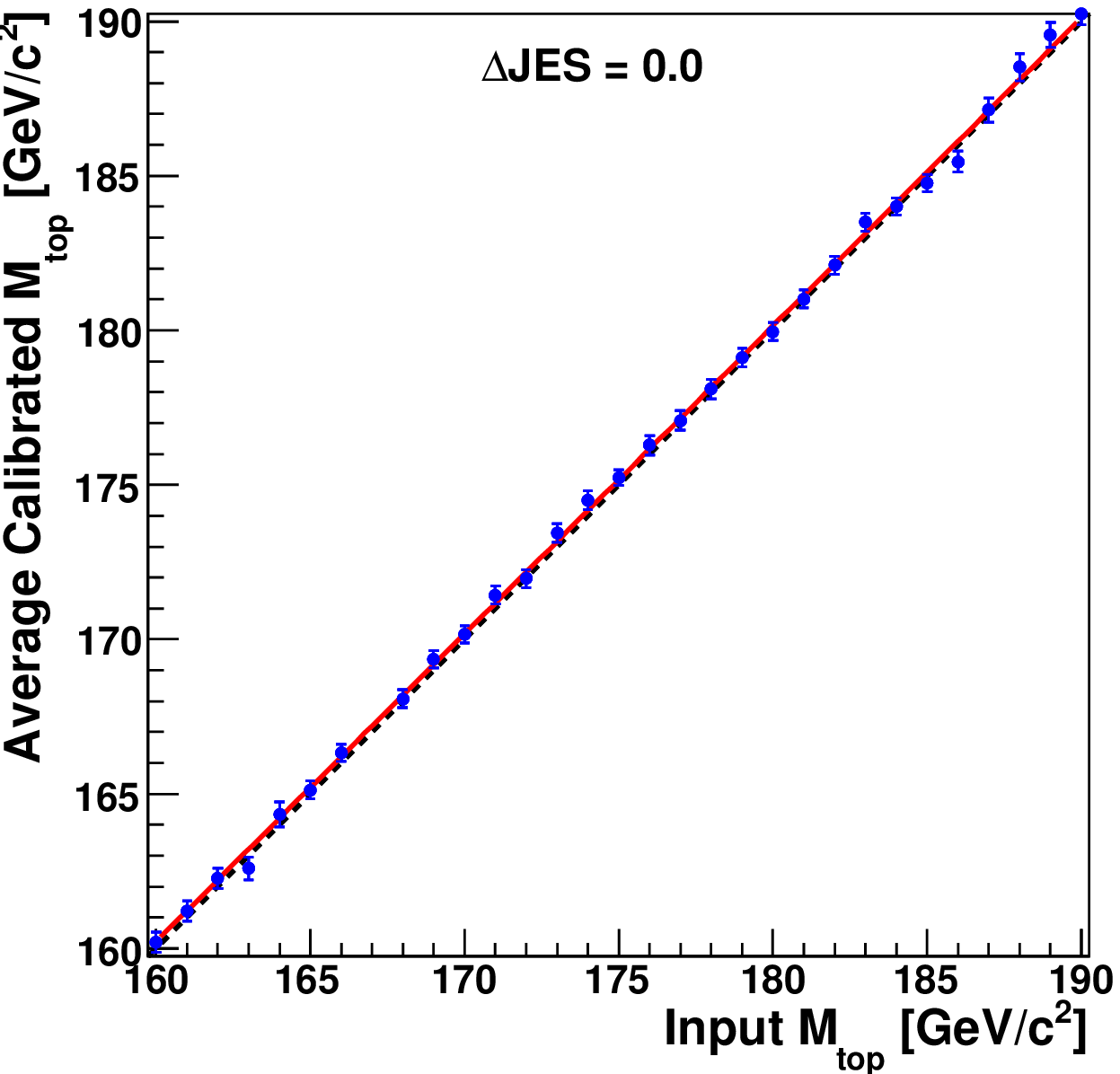,width=8cm}   \\ 
   \epsfig{file=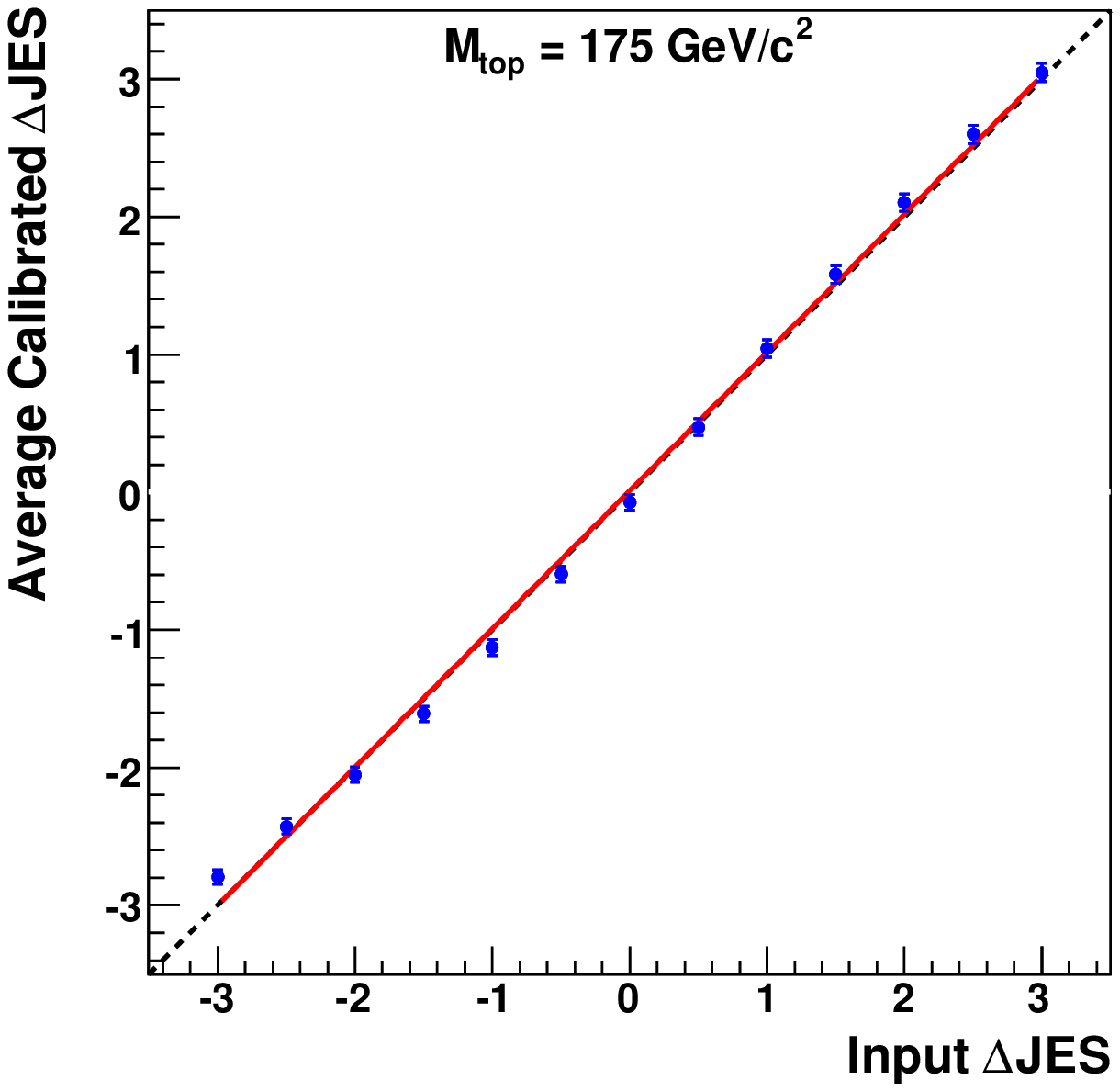,width=8cm} 

\end{tabular}
\caption{
         Examples of response linearity plots ($\langle M_{\mathrm{top}}^{\mathrm{corr}} \rangle$ vs $M_{\mathrm{top}}^{\mathrm{in}}$, 
         upper plot, and 
         $\langle \Delta\mathrm{JES}^{\mathrm{corr}} \rangle $ vs $\Delta\mathrm{JES}^{\mathrm{in}}$, lower plot) after the
         2D calibration.  The solid lines represent the linear functions
               which best fit the response as a function
               of the input values, while the dashed straight lines
               represent the ideal behavior.
        }
\label{fig:linmtopjes2}
\end{center}
\end{figure}

In Fig.~\ref{fig:linmtopjes2} examples of linearity plots
are shown for calibrated PEs. These plots, together with the numbers in Table~\ref{tab:calib}, 
show how the calibration removes any average bias.
To check that the uncertainties  $\delta M_{\mathrm{top}}^{\mathrm{corr}}$ 
and $\delta \Delta\mathrm{JES}^{\mathrm{corr}}$
are also unbiased, 
we consider the width of $M_{\mathrm{top}}^{\mathrm{corr}}$ 
and $\Delta\mathrm{JES}^{\mathrm{corr}}$ pull distributions, 
that is, the distributions of deviations of the calibrated values from 
the true inputs in the PEs, divided by the uncertainties themselves.
We find that the uncertainties are both underestimated, and multiplicative correction 
factors equal to $1.084$ for $\delta M_{\mathrm{top}}^{\mathrm{corr}}$ and to $1.115$ 
for $\delta \Delta \mathrm{JES}^{\mathrm{corr}}$ are needed. 
%
After these corrections the average expected uncertainty on the measured 
top quark mass and jet energy scale displacement 
for true $M_{\mathrm{top}}$ and $\Delta\mathrm{JES}$ around $175$\,GeV/$c^2$ and $0$ 
are
\begin{eqnarray}
                  \delta M_{\mathrm{top}}^{\mathrm{corr}}\,(\mathrm{stat} + \mathrm{JES})~~~ &  = & 2.0\,\mathrm{GeV}/c^{2}\,, \\ 
            \nonumber \\  
                  \delta \Delta\mathrm{JES}^{\mathrm{corr}}\,(\mathrm{stat} + M_{\mathrm{top}}) &  = & 0.45 \,.
\end{eqnarray}


\section{Systematic uncertainties}
\label{sec:syst}

The possible systematic uncertainties 
on the top quark mass and the jet energy scale measurements
have been studied
and are summarized in this section. 
These arise mostly from the measurement technique itself,
from uncertainties in the simulation of the $t \bar t$ events,
from mismodeling of the detector response, and from uncertainty on
the shapes of signal and background templates used to derive the p.d.f.'s and
to calibrate the measurement.
The uncertainties are usually evaluated by performing PEs, extracting pseudodata from
templates built using signal samples where the possible systematic effects 
have been considered and included.
Corrections to the shape of the raw background templates 
are performed as described in Sec.\,\ref{sec:bkgtemplates}
to obtain the corrected background templates corresponding 
to the effect one wants to study. 
On the contrary, nothing is changed in the
elements of the likelihood fit, because 
it is the default procedure that we want to apply to real
data and that, therefore, we have to test in case of possible mismodeling of the data themselves. 
The results from these PEs are then compared to the ones obtained by using default templates,
and the shifts in the average $M_{\mathrm{top}}^{\mathrm{corr}}$ and $\Delta\mathrm{JES}$$^{\mathrm{corr}}$ 
values are taken as
the estimate of the systematic uncertainties. 
In some cases the statistical uncertainties on the shifts may be larger than the shifts themselves, and
therefore we use conservatively the former as the systematic uncertainty.
In the following, after the description of each effect, we also quote in parentheses 
the values of the
corresponding uncertainties for the top quark mass and the jet energy scale respectively.
These values are then summarized in Table\,\ref{tab:obssyst}.

The 2D calibration removes the average biases, especially related to the
parametrization of the templates using smooth probability density functions.
Residual biases usually exist at single
($M_{\mathrm{top}}^{\mathrm{in}},\,\Delta\mathrm{JES}^{\mathrm{in}}$) points, and have to be taken into account.
We therefore consider the shift of the mean of the pull distributions with respect to 0
at each ($M_{\mathrm{top}}^{\mathrm{in}}$,\,$\Delta\mathrm{JES}^{\mathrm{in}}$) 
point to evaluate this {\em residual bias} systematic uncertainty, which, given the definition of
pull in Sec.~\ref{sec:sanity}, is defined as a 
function of the uncertainty on the calibrated measurements. To obtain the proper coverage of both
positive and negative biases we evaluate them separately, so that asymmetric uncertainties are 
finally considered. They are generally given by 
$(^{+0.37}_{-0.20}) \cdot \delta M_{\mathrm{top}}^{\mathrm{corr}}$
for $M_{\mathrm{top}}^{\mathrm{corr}}$ and 
 $(^{+0.43}_{-0.56}) \cdot \delta \Delta\mathrm{JES}^{\mathrm{corr}}$ 
for $\Delta\mathrm{JES}^{\mathrm{corr}}$. Specifying the values obtained 
in the measurement on the data, described in Sec.~\ref{sec:meas}, we obtain
$^{+0.8}_{-0.4}$\,GeV/$c^{2}$ on $M_{\mathrm{top}}^{\mathrm{corr}}$, $ ^{+0.18}_{-0.24} $ on $\Delta\mathrm{JES}^{\mathrm{corr}}$.
\par
The uncertainties on the parameters of the 2D calibration give a small uncertainty on the corrected values
$M_{\mathrm{top}}^{\mathrm{corr}}$ and $\Delta\mathrm{JES}^{\mathrm{corr}}$ which can be evaluated by the calibration functions
and the values of $M_{\mathrm{top}}$ and $\Delta\mathrm{JES}$ fitted in the data 
($ <0.1$\,GeV/$c^{2}$, $ <0.01 $).
\par
Many sources of systematic effects arise from uncertainties in modeling of
the hard interaction in simulated events. {\sc pythia} and {\sc herwig}\,\cite{Herwig} 
Monte Carlo generators differ in their hadronization  schemes and in their description of the 
underlying event and multiple interactions. The default signal samples 
have been generated with {\sc pythia} and therefore an uncertainty is obtained 
by using a sample generated using {\sc herwig}
($0.3$\,GeV/$c^{2}$, $0.25 $).
%
\par
Jets coming from possible emission of hard gluons 
might fall among the six leading jets and populate the tails in the top quark invariant mass distribution.
The amount of radiation from partons in the initial or final state
is set by parameters of the {\sc pythia} generator used to simulate signal events.
To study  these effects, templates are built using samples where
the values of the parameters have been changed with respect to the default, to increase or
to decrease the amount of radiation~\cite{massTMT}
($0.1$\,GeV/$c^{2}$, $0.06 $).
\par
Since the default jet energy corrections are derived from data samples 
deficient in heavy flavors~\cite{JESNIM}, 
an additional uncertainty  comes from considering the different properties 
of $b$ quarks.  We account for the uncertainties on the $b$-quark semileptonic branching ratios, 
fragmentation modeling, and calorimeter response to heavy-flavor hadrons
($0.2$\,GeV/$c^{2}$, $0.04 $).
\par
The different efficiency of the $b$-tagging algorithm 
on  data and  simulated events is usually considered by introducing
a constant scale factor ($b$-tag SF).
The overall uncertainty on this parameter affects the cross section
measurement described in Sec.\,\ref{sec:xsec}. 
However, such scale factor does not need to be considered regarding
the top quark mass measurement, because
it could slightly change only the population of the signal templates, but not their shape. 
On the other hand, variations of the latter could be caused by the possible dependence of 
the $b$-tag SF on the transverse energy of jets, which is then considered as a systematic effect
($0.1$\,GeV/$c^{2}$, $0.01 $).
\par
The uncertainty on the top quark mass coming from the likelihood fit
includes the uncertainty due to the jet energy scale. However,
as described in Sec.\,\ref{sec:presel}, this
uncertainty is the result of many independent effects with different
behavior with respect to properties of jets like $E_{T}$ and $\eta$\,\cite{JESNIM}, 
and therefore represents 
a leading-order correction. Second-order effects can arise from
uncertainties on the single corrections applied to the jet energies.
To evaluate these possible effects, we build signal templates by varying separately by $\pm 1\, \sigma$
the single corrections and, for each one of these variations, 
PEs were performed by using these templates and
not applying the
constraint ${\cal L}_{\Delta\mathrm{JES}_{\mathrm{constr}}}$ in the likelihood fit, 
as this term is related to effects of the full correction.
The resulting uncertainties have been added in quadrature to obtain a {\em residual} $\mathrm{JES}$
uncertainty on the top quark mass ($0.5$\,GeV/$c^{2}$).
\par
The choice of parton distribution functions (PDF) in the proton used
in Monte Carlo generators can affect 
the kinematics of simulated $t \bar t$ events
and thus the top quark mass measurement. We considered four sources of uncertainties\,: 
the difference arising from the use of the default CTEQ5L~\cite{CTEQ} PDF
and one calculated from the MRST group, MRST72~\cite{MRST};
the uncertainty depending on the value of $\alpha_{s}$, evaluated
by the difference between the use of MRST72 and MRST75 PDF's;
the uncertainty depending on the differences between the LO 
and NLO calculations of PDF'\,s, evaluated by the difference 
between using default CTEQ5L (LO) and CTEQ6M (NLO) PDF'\,s; 
and the uncertainties on PDF'\,s derived from experimental data uncertainties 
($^{+0.3}_{-0.2}$\,GeV/$c^{2}$, $^{+0.05}_{-0.04}$).
\par
The probability to have multiple $p\bar{p}$ interactions during the same bunch crossing 
is a function of the instantaneous luminosity. This is reflected in the
increasing number of primary vertices reconstructed in the events at higher luminosities.
We account for the fact that the simulated samples for the signal process 
do not model the actual luminosity profile of the data by considering the signal distributions
for events with 1, 2, 3, and $\ge 4$ reconstructed vertices separately. 
These distributions are then used
to obtain the templates by weighted averages, where the weights are evaluated as the
fractions of events with 1, 2, 3 and $\ge 4$ vertices observed in the data.
Moreover, a possible mismodeling of the dependence of the jet energy response 
as a function of the reconstructed number of primary vertices in simulated events is considered
($0.2$\,GeV/$c^{2}$, $0.01 $). 
\par
Uncertainties from modeling of color reconnection 
               effects~\cite{CR} are estimated by comparing
               the results of two sets of PEs performed drawing
               pseudodata from templates built using two different
               samples of events simulated by {\sc pythia}.
               The samples are generated with two different
               sets of parameters, corresponding to two different 
               models of color reconnection ($0.4$\,GeV/$c^{2}$, $0.08 $).
\par
The shapes of the signal and background distributions are affected 
by uncertainties due to the limited statistics of the simulated events 
and data 
samples used to build them. These uncertainties affect the results of a measurement which is 
performed maximizing an 
unbinned likelihood where parametrized p.d.f.'\,s, fitted to default templates, are evaluated. 
We address this effect obtaining 100 sets of templates by statistical fluctuations of default ones and
performing pseudoexperiments drawing data from each of these sets separately.
From each set we obtain an average value for  $M_{\mathrm{top}}^{\mathrm{corr}}$ 
and $\Delta\mathrm{JES}^{\mathrm{corr}}$,
and the spread of these values is taken as the systematic uncertainty
($0.3$\,GeV/$c^{2}$, $0.07 $).
\par
Besides the purely statistical effects, quoted above, the shape of the 
background templates also has  uncertainties 
due to the corrections for the presence of signal events in the pretag sample, Sec.~\ref{sec:bkgtemplates},
and to the
systematic uncertainty on the background normalization, Sec.~\ref{sec:cutoptimization}.  
We address this source of systematic uncertainty by the same technique used for the 
statistical contributions, that 
is by obtaining a set of 100 background templates where the content of each bin is separately 
fluctuated by Gaussian distributions centered on the default bin content and with a 
width equal to its systematic uncertainty, and taking the spread of results from PEs
as the systematic uncertainty
($0.1$\,GeV/$c^{2}$, $0.02 $).
\par
%

Table~\ref{tab:obssyst} shows a summary of all the systematic uncertainties and their 
sum in quadrature, which gives a total systematic uncertainty of 
$^{+1.2}_{-1.0}$\,GeV/$c^2$ for the $M_{\mathrm{top}}$ measurement and 
$^{+0.34}_{-0.37}$ for the $\Delta\mathrm{JES}$. 

\begin{table}[hbt!]
\caption[c 2]{
              Systematic uncertainties 
	      and their sizes for the top-quark mass and the jet
              energy scale measurements. 
              The total uncertainty is obtained by the sum in quadrature of single contributions. 
             }
\label{tab:obssyst}
\begin{tabular}{lcc}
\hline
\hline
Source                           & \parbox[l][5mm][c]{17mm}{$\delta M_{\mathrm{top}}^{syst}$} &  $\delta \Delta \mathrm{JES}^{syst}$ \\
                                 &                              (GeV/$c^2$)          &                                      \\
\hline

Residual bias                          &   \parbox[c][7mm][c]{17mm}{ $ ^{+0.8}_{-0.4}$}   & $ ^{+0.18}_{-0.24} $   \\
2D calibration                         &   \parbox[c][7mm][c]{17mm}{ $ <\,0.1 $}          & $ <\, 0.01 $   \\
Generator                              &   \parbox[c][5mm][c]{17mm}{ $ 0.3 $}             & $ 0.25 $   \\ 
ISR/FSR                                &   \parbox[c][5mm][c]{17mm}{ $ 0.1 $}             & $ 0.06 $   \\ 
$b$-jet energy scale                   &   \parbox[c][5mm][c]{17mm}{ $ 0.2 $}             & $ 0.04 $   \\ 
$b$-tag SF $E_T$ dependence                    &   \parbox[c][5mm][c]{17mm}{ $ 0.1 $}             & $ 0.01 $   \\ 
Residual $\mathrm{JES}$                &   \parbox[c][5mm][c]{17mm}{ $ 0.5 $}             & ...   \\
PDF                                    &   \parbox[c][7mm][c]{17mm}{ $ ^{+0.3}_{-0.2}$}   & $ ^{+0.05}_{-0.04} $   \\
Multiple $p\bar{p}$ interactions ~~~      &   \parbox[c][5mm][c]{17mm}{ $ 0.2 $}             & $ 0.01 $   \\
Color reconnection                     &   \parbox[c][5mm][c]{17mm}{ $ 0.4 $}             & $ 0.08 $   \\
Statistics of templates                &   \parbox[c][5mm][c]{17mm}{ $ 0.3 $}             & $ 0.07 $   \\
Background shape                       &   \parbox[c][5mm][c]{17mm}{ $ 0.1 $}             & $ 0.02 $   \\ 

\hline
Total                                  &   \parbox[c][7mm][c]{17mm}{$ ^{+1.2}_{-1.0} $} & $  ^{+0.34}_{-0.37}$   \\ 
\hline
\hline
\end{tabular}
\centering
\end{table}


\section{\boldmath Top Mass and $\mathrm{JES}$ measurements}
{\label{sec:meas}

After the kinematical selection with $N_{\mathrm{out}}\ge 0.90$ ($\ge 0.88$) and  
$\chi^2 \le 6$ ($\le 5$), we are left with 3452 (441) events with one ($\ge 2$) tag(s). 
The background amounts to $2785 \pm 83$ ($201 \pm 29$) for events with one ($\ge 2$) tag(s). 

For these events a top quark mass has been reconstructed using the likelihood fit
described in Sec.~\ref{sec:like} and applied to the data sample. Once the calibration procedure
and corrections are applied, as described in Sec.~\ref{sec:sanity},
the best estimate of the top quark mass is 
\begin{eqnarray}
M_{\mathrm{top}} = 174.8 \pm 2.4\, (\mathrm{stat}+ \mathrm{JES})\,\mathrm{GeV}/c^{2} \,, 
\end{eqnarray}
while the value obtained for the jet energy scale displacement is 
\begin{eqnarray}
\Delta\mathrm{JES} = -0.30 \pm 0.47\, (\mathrm{stat}+M_{\mathrm{top}}) \,. 
\end{eqnarray}
We can also evaluate separately the purely statistical contributions obtaining
\begin{eqnarray}
M_{\mathrm{top}} = 174.8 \pm 1.7\,(\mathrm{stat}) \pm 1.6\,(\mathrm{JES})\,\mathrm{GeV}/c^{2} \,, 
\end{eqnarray}
and
\begin{eqnarray}
\Delta\mathrm{JES} = -0.30 \pm 0.35\,(\mathrm{stat}) \pm 0.32\,(M_{\mathrm{top}}) \,. 
\end{eqnarray}

The plot in Fig.\,\ref{fig:contours} shows the measured values together with 
the log-likelihood contours corresponding to 1\,$\sigma$, 2\,$\sigma$, and 3\,$\sigma$ uncertainty 
on the value of the top quark mass\,\cite{pdg}. The slope of the 
major axis of the contours denotes that the measurements of $M_{\mathrm{top}}$ and $\Delta\mathrm{JES}$
have a negative correlation, and the value of the correlation coefficient 
obtained from the likelihood fit is $-0.68$.

\begin{figure}[htbp!]
\begin{center}
%
     \begin{tabular}{c}
               \epsfig{file=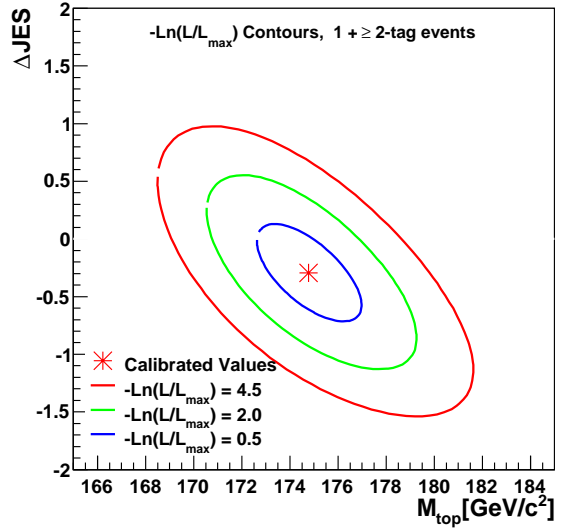,width=8cm}
         \end{tabular}
\caption{
         Negative log-likelihood contours for the likelihood fit performed
         for the $M_{\mathrm{top}}$ and $\Delta\mathrm{JES}$ measurements.
	 The minimum is also shown and corresponds to the values measured in the data.
         The contours are drawn at values of 0.5, 2.0, and 4.5 of the increase of
         the log-likelihood from the minimum value. These curves correspond
         to 1\,$\sigma$, 2\,$\sigma$, and 3\,$\sigma$ uncertainty on the measurement of the top
         quark mass.
        }	 

 \label{fig:contours}
\end{center}
\end{figure}

The plots in Fig.~\ref{fig:topmeas} show the $m_{t}^{\mathrm{rec}}$ 
distributions 
for the data compared to the expected background and the signal for a top quark mass 
of $175.0$\,GeV/$c^2$ and a jet energy scale displacement of $-0.3$, 
that is, the values of simulated $M_{\mathrm{top}}$ and $\Delta\mathrm{JES}$ as close as possible to the measurements
in the data.  The signal and background contributions are normalized to the respective  
number of events as fitted in the data.

\begin{figure}[htbp!]
\begin{center}
          \begin{tabular}{c}
	      \epsfig{file=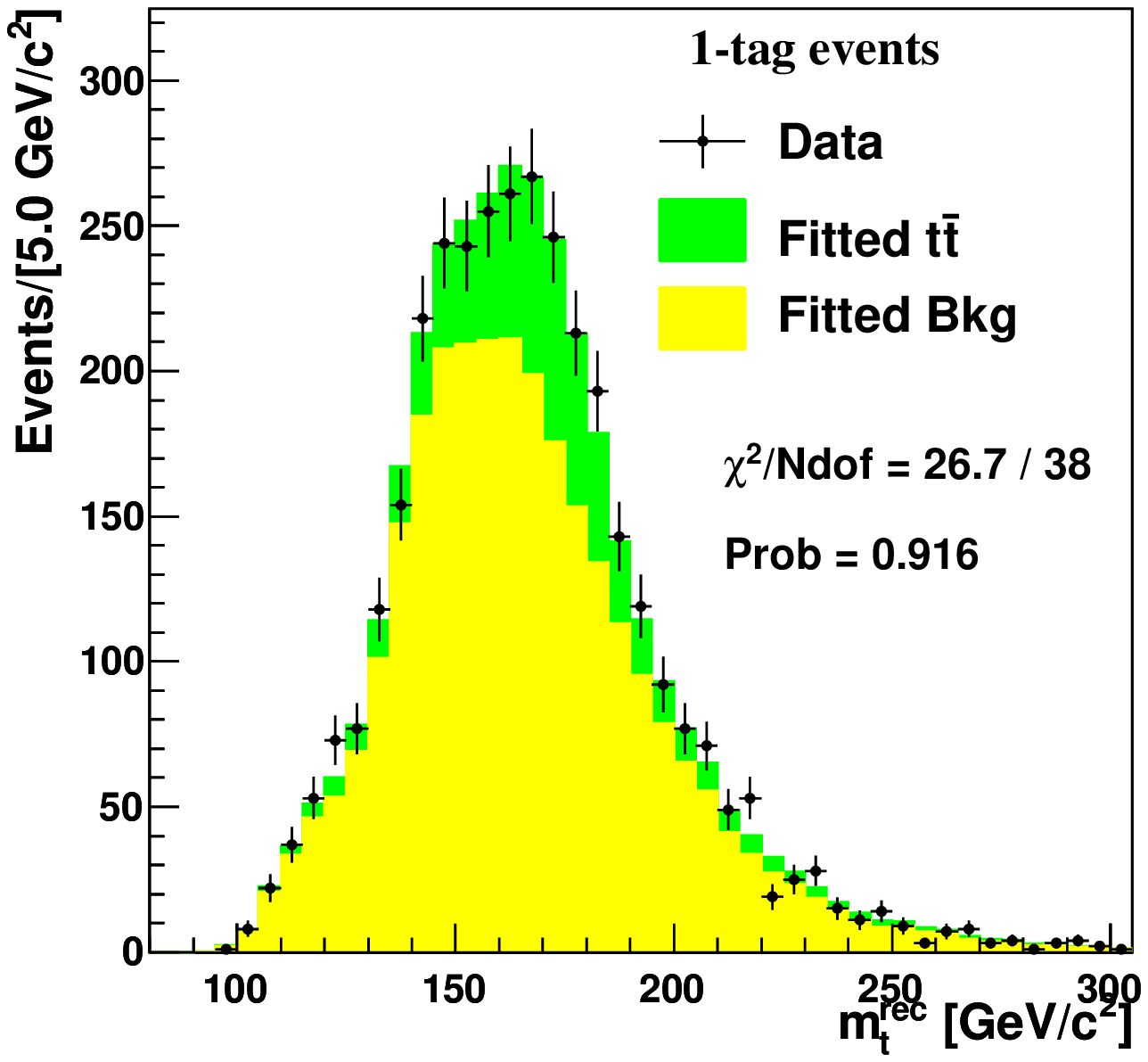,width=8cm}  \\
              \epsfig{file=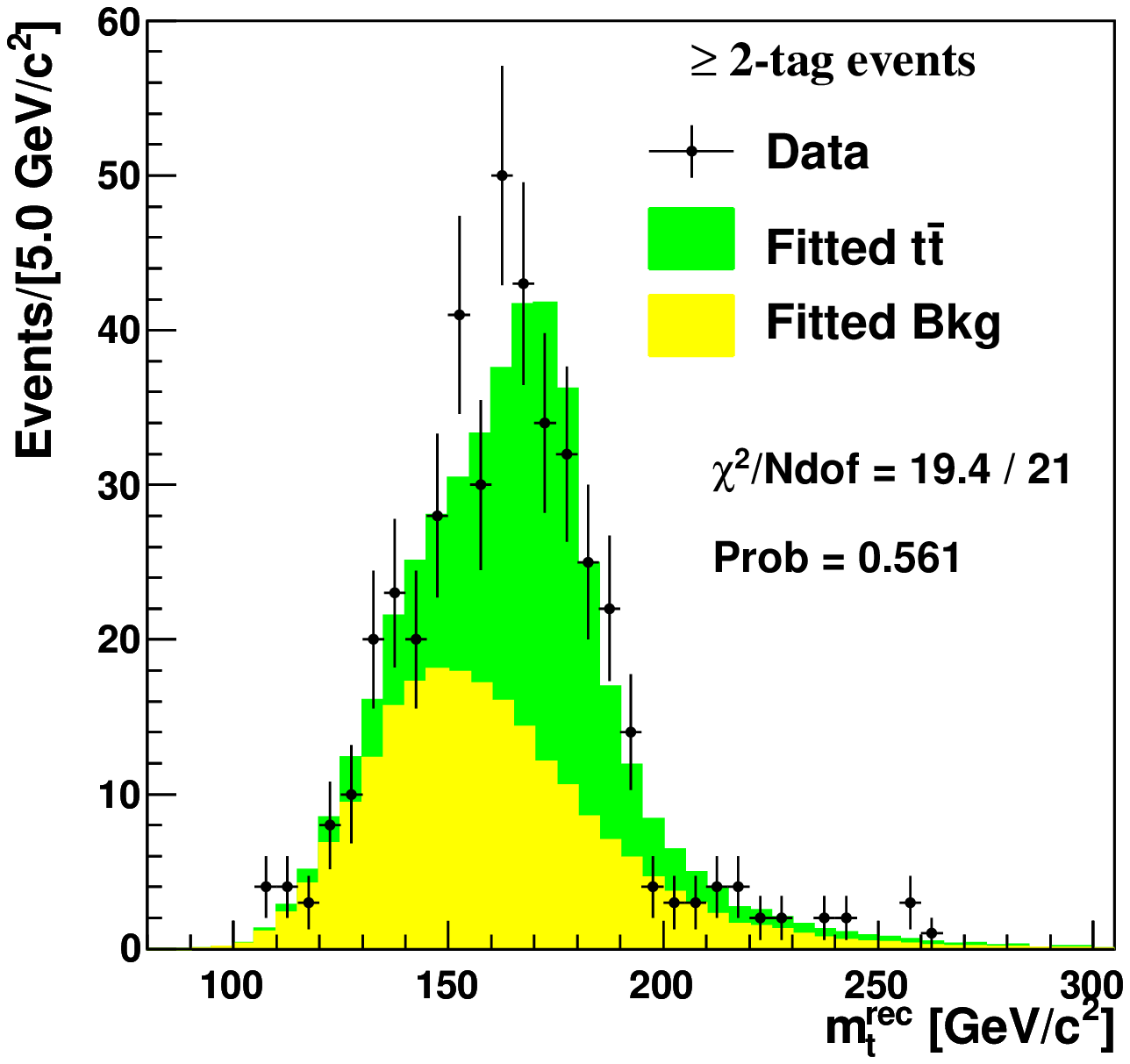,width=8cm} 
                                            
          \end{tabular}
\caption{Histograms of $m_{t}^{\mathrm{rec}}$ 
          as obtained in the data (black points) for 1-tag (upper plot) and $\ge 2$-tag events (lower plot)  
         are compared to the distributions from signal and background corresponding to 
         $M_{\mathrm{top}}= 175$~GeV/$c^{2}$ and $\Delta\mathrm{JES} = -0.3$. 
          The expected histograms are normalized to the measured values for the average number of signal 
          and background events. The values of the purely statistical
          $\chi^{2}$ and of its probability are reported on each plot, showing the overall
          agreement between the data and the distributions corresponding to the fitted values
          of $M_{\mathrm{top}}$ and $\Delta\mathrm{JES}$. 
        }
\label{fig:topmeas}
\end{center}
\end{figure}

 The plots in Fig.~\ref{fig:sigma} compare the measured statistical uncertainty, 
just after the 2D calibration, 
with the expected distribution from default pseudoexperiments 
using as inputs $M_{\mathrm{top}} = 175.0$\,GeV/$c^{2}$ 
and $\Delta\mathrm{JES} = -0.3$. We find that the probability of achieving a better 
sensitivity is $91.6\%$ for $M_{\mathrm{top}}$ and $81.2\%$ for $\Delta\mathrm{JES}$.

\begin{figure}[htbp!]
\begin{center}
          \begin{tabular}{c}
                             \epsfig{file=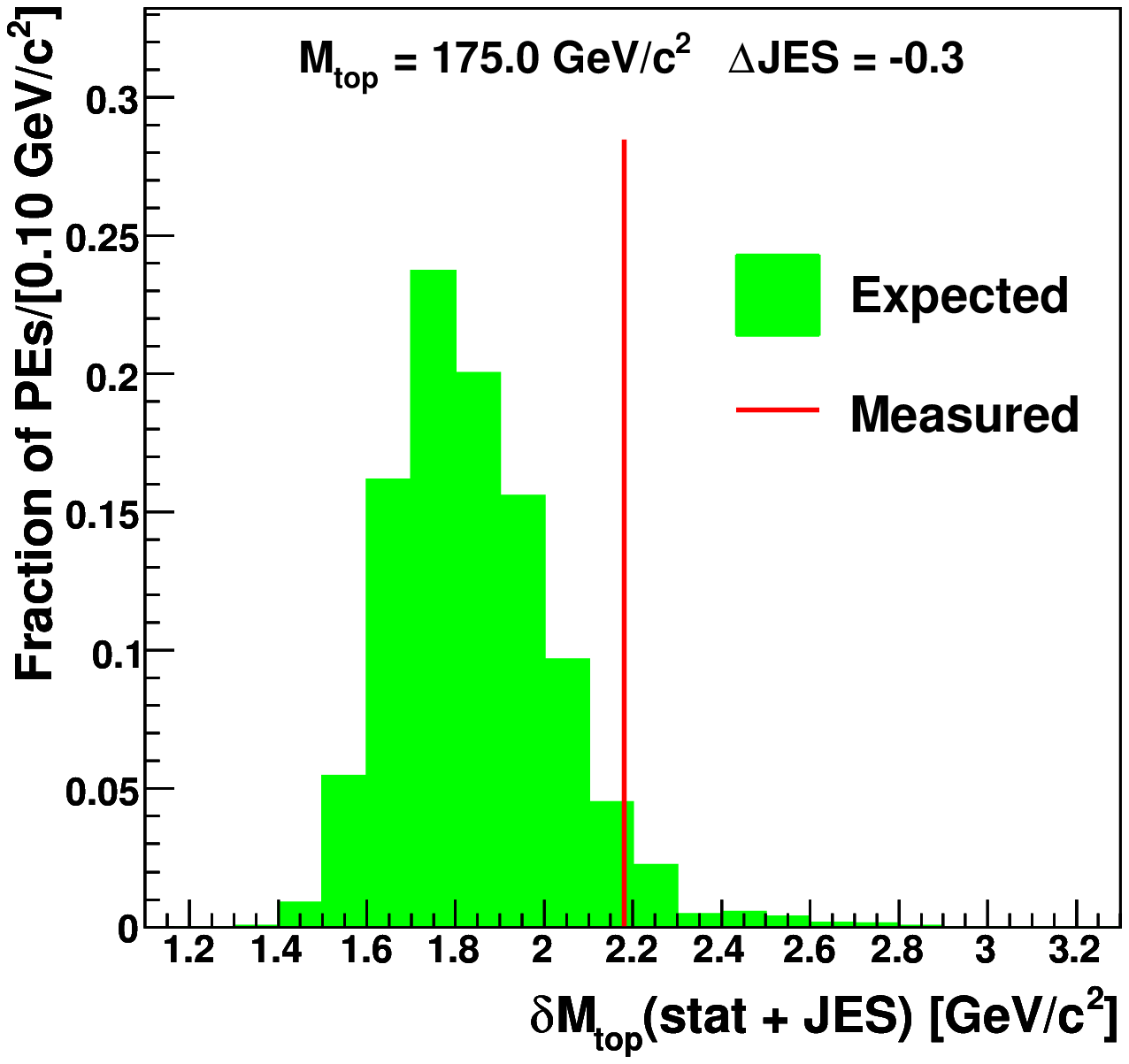,width=8cm} \\
                             \epsfig{file=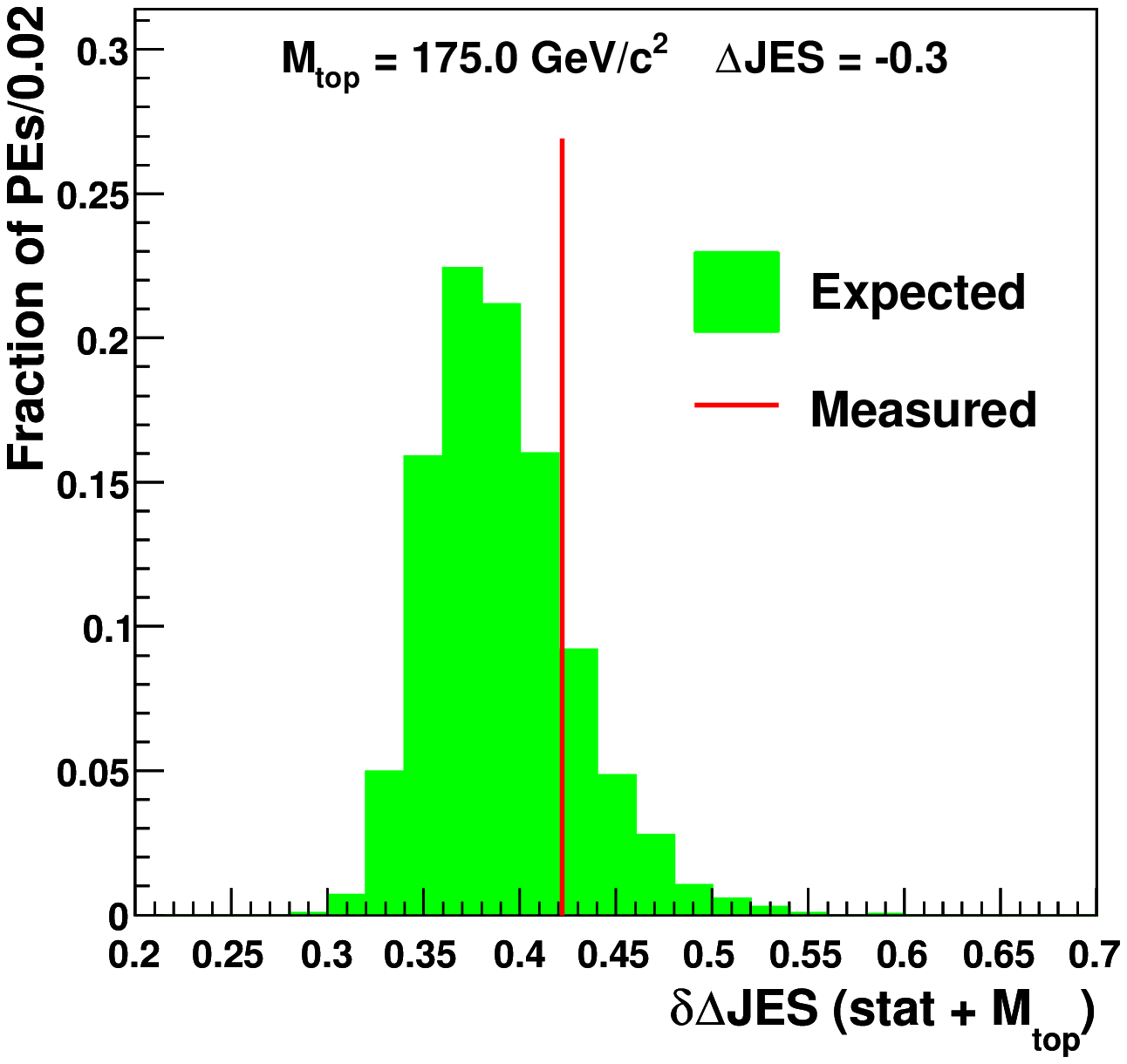,width=8cm}                             
          \end{tabular}
\caption{Distributions of the uncertainties on the top quark mass (upper plot) and the 
         jet energy scale displacement 
         (lower plot)
         as expected from default PEs performed using as input values $M_{\mathrm{top}}^{\mathrm{in}} = 175.0$\,GeV/$c^{2}$ 
         and $\Delta\mathrm{JES}^{\mathrm{in}} = -0.3$.
         The vertical lines indicate the uncertainties obtained in our reported
         result.
        }
\label{fig:sigma}
\end{center}
\end{figure}

Summarizing, the measured values for the top quark mass and the jet energy scale are
\begin{eqnarray}
 M_{\mathrm{top}} =  174.8 \, \pm 2.4\,(\mathrm{stat} + \mathrm{JES})\, ^{+1.2}_{-1.0}\,(\mathrm{syst})\,\mathrm{GeV}/c^{2}\,,~~    
\end{eqnarray}
\begin{eqnarray}
  \Delta \mathrm{JES} =  -0.30\, \pm 0.47\,(\mathrm{stat} + M_{\mathrm{top}})\,^{+0.34}_{-0.37}\,(\mathrm{syst}) \,,~~ 
\end{eqnarray}
which, isolating the purely statistical contributions and
adding the uncertainties from $\mathrm{JES}$ and $M_{\mathrm{top}}$ to the respective 
systematic uncertainties, can also be written as
\begin{eqnarray}
 M_{\mathrm{top}} =  174.8~\pm 1.7\,(\mathrm{stat})~ ^{+2.0}_{-1.9}\,(\mathrm{syst})~\mathrm{GeV}/c^{2}\,,~~ 
\end{eqnarray}
\begin{eqnarray}
  \Delta \mathrm{JES} =  -0.30~ \pm 0.35\,(\mathrm{stat})~ ^{+0.47}_{-0.49}\,(\mathrm{syst}) \,.~~  
\end{eqnarray}

This measurement of the top quark mass has been used in the current world average\,\cite{latestCDFD0mass}.
%
%
%
\section{Cross section measurement}
\label{sec:xsec}

The procedure used to measure the top quark mass also returns the average number 
of signal events expected, given the selected data samples.
These results can be turned into a measurement of the $t\bar t$ cross section,
as follows.


\subsection{The likelihood function}
\label{sec:sigmalike}

From the number of signal events, $n_{s}^{1\,\mathrm{tag}}$ and $n_{s}^{\ge 2 \,\mathrm{tags}}$, as obtained
from the mass likelihood fit, we derive a measurement
of the $t\bar{t}$ production cross section considering the efficiency for selecting
a $t\bar{t}$ event in the two tagging categories.

The cross section measurement is performed by maximizing a likelihood function which can be
divided into two parts\,:
\begin{eqnarray}
%
%
 {\cal L} = {\cal L}_{1\,\mathrm{tag}} \times {\cal L}_{\ge 2\,\mathrm{tags}}\,, 
\end{eqnarray} 
where 
each term can be expressed as\,:
\begin{eqnarray}
 {\cal L}_{1,\ge 2\, \mathrm{tags}} = {\cal L}_{\sigma_{t\bar{t}}} \times {\cal L}_{\epsilon}\,,  
\end{eqnarray}
where
\begin{eqnarray}
{\cal L}_{\sigma_{t\bar{t}}} & =  & 
\exp{\left[ - ~ \frac{ \left(\sigma_{t\bar{t}} \cdot \epsilon \cdot {\mathrm L} - n_{s}\right)^{2} }{2\sigma_{n_{s}}^{2}}\right]} 
\end{eqnarray}
contains all the parameters of the fit, i.e. the production cross section $\sigma_{t\bar{t}}$,
the integrated luminosity ${\mathrm L}$, the signal efficiency $\epsilon$, and  
the signal yield $n_{s} \pm \sigma_{n_{s}}$, as given by the mass measurement, 
while ${\cal L}_{\epsilon}$ is a Gaussian term constraining the  efficiency within
its statistical uncertainty.

The efficiencies are evaluated using a sample of about $4 \times 10^{6}$  $t\bar{t}$ events 
generated with \mbox{$M_{\mathrm{top}} = 175$\,GeV/$c^{2}$} and assuming  \mbox{$\Delta\mathrm{JES}= -0.3$},
i.e. the value we measured by the mass likelihood fit,
and are summarized along with signal yields and other parameters in Table\,\ref{tab:xsecpar}.

While studying the performance of the procedure, using pseudoexperiments produced assuming a given 
input cross section, we observe the need to introduce a small correction. 
The outcome of the fit needs
to be multiplied by a factor $k_\sigma=0.982\pm 0.008$ in order to obtain an unbiased measurement
of the cross section.
\begin{table}
\begin{center}
 \caption{Input variables to the cross section evaluation. 
          For the signal yields, the first uncertainty is the purely statistical one.}
 \label{tab:xsecpar}
   \begin{tabular}{lc}
     \hline\hline
     Variable~~~~~~~~~~~~~~~~                 &  \parbox[c][5mm][c]{30mm}{ Input value }            \\
     \hline
     Signal yield, one tag             &  \parbox[c][5mm][c]{30mm}{$ 643 \pm 59 \pm 54$ }  \\
     Signal yield, $\ge 2$ tags~~~~~~~~~~~      &  \parbox[c][5mm][c]{30mm}{$ 216 \pm 21 \pm 14$ }  \\
     Efficiency, one tag               &  \parbox[c][5mm][c]{30mm}{$ (3.55 \pm 0.01)\%$ }        \\ 
     Efficiency, $\ge 2$ tags        &  \parbox[c][5mm][c]{30mm}{$ (1.00 \pm 0.01)\%$ }        \\ 
     Integrated luminosity           &  \parbox[c][5mm][c]{30mm}{$ 2874 \pm 172$\,pb$^{-1}$ }  \\
     \hline\hline

    \end{tabular}
 \end{center}

\end{table}

From the maximization of the likelihood, we obtain a central value for the $t\bar{t}$ production cross
section
\begin{eqnarray}
   \sigma_{t\bar{t}} =7.2 \pm 0.5 (\textnormal{stat})\pm 0.4(\textnormal{lum})\,\mathrm{pb}\,, 
\end{eqnarray}
evaluated assuming $M_{\mathrm{top}}=175$ GeV/c$^2$ and $\Delta\mathrm{JES}=-0.3$, close to the values measured in Sec.\,\ref{sec:meas}.
The first uncertainty is the statistical one, while the second one derives from the 6\% uncertainty 
on the integrated luminosity.
As the signal efficiencies depend strongly on the assumed values for  $M_{\mathrm{top}}$ and $\Delta\mathrm{JES}$, 
the measured $t\bar t$ cross section also has the same dependence. 
For reference we report in Table\,\ref{tab:xsec-mass-jes} the cross 
sections corresponding to other $(M_{\mathrm{top}},\Delta\mathrm{JES})$ points with a top quark mass 
near the current CDF average. 
In this case we assume $\Delta\mathrm{JES}=0$ and the systematic uncertainty on JES 
is increased from 6.1\% to 9.2\%, corresponding to 
changing the $\Delta$JES by $\pm 1$ rather than by $\pm 0.6$ units, that is the sum in quadrature of
the uncertainties on the measured jet energy scale, Sec.\,\ref{sec:meas}.

\begin{table}
\begin{center}
 \caption{Cross section as evaluated assuming different values for $M_{\mathrm{top}}$ and $\Delta\mathrm{JES}$.}
 \label{tab:xsec-mass-jes}
   \begin{tabular}{ccc}
     \hline\hline
\parbox[c][5mm][c]{30mm}{$M_{\mathrm{top}}$ (GeV/c$^2$)}  &  ~~~~~~$\Delta\mathrm{JES}$~~~~~~ & ~~~~~~$\sigma_{t\bar t}$ (pb) \\
\hline
\parbox[c][5mm][c]{30mm}{175.0} & -0.3 & 7.24 \\
\parbox[c][5mm][c]{30mm}{175.0} &  0.0 & 7.00 \\
\parbox[c][5mm][c]{30mm}{172.5} &  0.0 & 7.21 \\
\parbox[c][5mm][c]{30mm}{170.0} &  0.0 & 7.29 \\

     \hline\hline

    \end{tabular}
 \end{center}

\end{table}

\subsection{Systematic uncertainties}
\label{sec:sigmasyst}

Most of the sources of systematic uncertainties affecting
                the measurement of $\sigma_{t\bar{t}}$ are the same
                discussed for the measurement of the top quark mass.
                We just need to evaluate their effects both on the signal 
                yields and on the signal efficiencies in order to derive 
                the effects on the cross section.
There are few other sources of systematic uncertainty specific to a cross section measurement which have 
not been discussed in Sec.\,\ref{sec:syst}, because they
affect only the signal efficiencies.
These include the uncertainty on the calibration constant, $k_\sigma$, 
on the $W \to \mathrm{hadrons} $ branching ratio (BR)~\cite{pdg},
on the trigger simulation and on the distribution of the primary vertex $z$-coordinate.
As for the effect of the JES uncertainty on the efficiency, we have evaluated it by changing the $\Delta$JES by
$\pm 0.6$ units with respect to the measured value $\Delta\mathrm{JES}=-0.3$. 
Residual effects due to individual 
levels of corrections have been accounted for, too.
The relative uncertainties $\Delta \sigma_{t\bar{t}} / \sigma_{t\bar{t}} $ for the individual
sources are summarized in Table\,\ref{tab:xsecsyst}.
Considering their sum in quadrature, the $t\bar{t}$ production cross section amounts to 
\begin{eqnarray}
  \sigma_{t\bar t} = 7.2 \, \pm \, 0.5({\rm stat}) \, \pm \, 1.0({\rm syst}) \, \pm \, 0.4({\rm lum})\,\mathrm{pb}\,,~~
%
\end{eqnarray}
assuming  $M_{\mathrm{top}} = 175$\,GeV/$c^{2}$ and $\Delta\mathrm{JES}=-0.3$.

\begin{table}
\begin{center}
 \caption{ 
           Systematic uncertainties 
           and their relative sizes
           for the cross section measurement. 
           The total uncertainty is obtained by the sum in quadrature of single contributions. 
         }
 \label{tab:xsecsyst}
   \begin{tabular}{lc}
     \hline\hline

Source & \parbox[c][5mm][c]{25mm}{$\Delta \sigma_{t\bar{t}} / \sigma_{t\bar{t}} $ (\%)}\\
\hline

Calibration factor                  & \parbox[c][5mm][c]{25mm}{   0.8 } \\
Generator                           & \parbox[c][5mm][c]{25mm}{   4.2 } \\
ISR/FSR                             & \parbox[c][5mm][c]{25mm}{   0.6 } \\
$b$-jet energy scale                & \parbox[c][5mm][c]{25mm}{   2.8 } \\
$b$-tag SF $E_T$ dependence         & \parbox[c][5mm][c]{25mm}{   5.4 } \\
PDF                                 & \parbox[c][5mm][c]{25mm}{   3.4 } \\
Multiple $p\bar{p}$ interactions~~~~~~~~~  & \parbox[c][5mm][c]{25mm}{   2.5 } \\
Color reconnection                  & \parbox[c][5mm][c]{25mm}{   0.8 } \\
Templates statistics                & \parbox[c][5mm][c]{25mm}{   0.8 } \\
Background shape                    & \parbox[c][5mm][c]{25mm}{   0.3 } \\
Background normalization            & \parbox[c][5mm][c]{25mm}{   8.2 } \\
$\mathrm{JES}$                      & \parbox[c][5mm][c]{25mm}{   6.1 } \\
Residual $\mathrm{JES}$             & \parbox[c][5mm][c]{25mm}{   2.1 } \\
Primary vertex                      & \parbox[c][5mm][c]{25mm}{   0.2 } \\
BR($W\to \mathrm{hadrons}$)         & \parbox[c][5mm][c]{25mm}{   0.8 } \\
Trigger                             & \parbox[c][5mm][c]{25mm}{   1.8 } \\

\hline
Total                               & \parbox[c][5mm][c]{25mm}{  13.7 } \\      
     \hline\hline

    \end{tabular}
 \end{center}

\end{table}


\section{Conclusions}
\label{sec:fine}

Using a very effective neural-network-based kinematical selection and a 
$b$ jet identification technique, 
we measure  the top quark mass to be 
\begin{eqnarray}
 M_{\mathrm{top}}=174.8\, \pm \, 2.4({\rm stat +JES}) \,^{+1.2}_{-1.0}({\rm syst})\,\mathrm{GeV}/c^{2}\,,~~ 
\end{eqnarray} \\
and the $t\bar t$ production cross section to be 
\begin{eqnarray}
  \sigma_{t\bar t} = 7.2 \, \pm \, 0.5({\rm stat}) \, \pm \, 1.0({\rm syst}) \, \pm \, 0.4({\rm lum})\,\mathrm{pb}\,.~~
%
\end{eqnarray}

These values represent  the most precise measurements to date of $M_{\mathrm{top}}$ 
and $\sigma_{t\bar t}$ in 
the all-hadronic decay channel.
The results are consistent with the measurements obtained
                in other decay channels by CDF\,and D{0} 
                Collaborations\,\cite{latestCDFD0mass, latestCDFD0xsec} 
                and, as it concerns $\sigma_{t\bar t}$, 
                with the theoretical predictions evaluated at the
                value of the top quark mass obtained in our
                measurement\,\cite{theo-xsec}.


\bigskip
\section{Acknowledgments} 

We thank the Fermilab staff and the technical staffs of the participating institutions 
for their vital contributions. 
This work was supported by the U.S. Department of Energy and National Science Foundation; 
the Italian Istituto Nazionale di Fisica Nucleare; 
the Ministry of Education, Culture, Sports, Science and Technology of Japan; 
the Natural Sciences and Engineering Research Council of Canada; 
the National Science Council of the Republic of China; 
the Swiss National Science Foundation; the A.P. Sloan Foundation; 
the Bundesministerium f\"ur Bildung und Forschung, Germany; 
the World Class University Program, the National Research Foundation of Korea; 
the Science and Technology Facilities Council and the Royal Society, United Kingdom; 
the Institut National de Physique Nucleaire et Physique des Particules/CNRS; 
the Russian Foundation for Basic Research; 
the Ministerio de Ciencia e Innovaci\'{o}n, and Programa Consolider-Ingenio 2010, Spain; 
the Slovak R\&D Agency; and the Academy of Finland.


%% file: Paper_Final_rev_mar10.bbl
\begin{thebibliography}{99}
\bibitem{met} The missing transverse energy $\vec{\met}$ is calculated as the vector sum 
              of the energy 
              in each calorimeter tower multiplied by a unit vector in the azimuthal 
              direction of the tower. If isolated
              high momentum muons are found in the event, the $\vec{\met}$ 
              is corrected by subtracting the
              muon energy in the calorimeter and adding the muon $p_T$ to 
              the vector sum.  $\met$ is defined as 
              the magnitude of  $\vec{\met}$.
\bibitem{twotagfit} The $\ge 2$-tag sample actually consists of events with 2 or 3 
                    tagged jets. 
                    When 3 tags are present, the two leading ones are assigned to $b$ quarks.
\end{thebibliography}
